\documentclass[superscriptaddress, groupedaddress, twocolumn, amsmath,amssymb, aps, prl]{revtex4}
\usepackage{graphicx}
\usepackage{bm}
\begin{document}

\title{Spin-orbital-angular-momentum coupled Bose-Einstein condensates}

\author{H.~-R. Chen}
\author{K.~-Y. Lin}
\author{P.~-K. Chen}
\author{N.~-C. Chiu}
\author{J.~-B. Wang}
\author{C.~-A. Chen}
\author{P.~-P. Huang}
\affiliation{Institute of Atomic and Molecular Sciences, Academia Sinica, Taipei, Taiwan 10617}
\author{S.~-K. Yip}
\affiliation{Institute of Atomic and Molecular Sciences, Academia Sinica, Taipei, Taiwan 10617}
\affiliation{Institute of Physics,Academia Sinica, Taipei, Taiwan 11529}
\author{Yuki Kawaguchi}
\affiliation{Department of Applied Physics, Nagoya University, Nagoya, 464-8603, Japan}
\author{Y.~-J. Lin}
\email{linyj@gate.sinica.edu.tw}
\affiliation{Institute of Atomic and Molecular Sciences, Academia Sinica, Taipei, Taiwan 10617}

\date{\today}

\begin{abstract}
We demonstrate coupling between the atomic spin and
orbital-angular-momentum (OAM) of the atom's center-of-mass motion
in a Bose-Einstein condensate (BEC). The coupling is induced by
Raman-dressing lasers with a Laguerre-Gaussian beam, and creates
coreless vortices in a $F=1$ $^{87}$Rb spinor BEC. We observe
correlations between spin and OAM in the dressed state and
characterize the spin texture; the result is in good agreement with
the theory. In the presence of the Raman field our dressed state is
stable for 0.1~s or longer, and it decays due to collision-induced
relaxation. As we turn off the Raman beams, the vortex cores in the
bare spin $|m_F=1\rangle$ and $|-1\rangle$ split. These spin-OAM
coupled systems with the Raman-dressing approach have great
potential for exploring new topological textures and quantum states.
\end{abstract}

\maketitle

One of the most exciting and productive research directions for
ultracold atoms has been engineering interesting Hamiltonians for
creating atomic gas analogs to iconic condensed-matter
models~\cite{Bloch2008}. Another motivation is to create systems
with fundamentally new regimes of quantum, topological, or other
forms of matter with no analogs elsewhere~\cite{Galitski2013}. A
landmark of this theme was the creation of synthetic gauge
potentials that act upon neutral atoms as if they were charged
particles~\cite{lin09,aidelsburger2013,miyake2013,Struck2012,Parker2013,jotzu2014,Dalibard11,Goldman2013,Zhai2015}.
A key experimental technique for producing such synthetic gauge
fields is optical Raman coupling between different internal spin
states where photon momentum is transferred to the atoms as the spin
state changes. This technique leads naturally to the kind of
``spin-orbit coupling" (SOC) seen in solids where the linear
momentum of electrons (atoms in our analog system) is coupled to
their spin: $\hbar\vec{k_e}\cdot \vec{s}$.  We refer to this as
spin-linear-momentum coupling
(SLMC)~\cite{lin11,Wu2016,Huang2016,Zhai2015}.
%; while a general SOC for neutral atoms is coupling their spin to the external degree of freedom.

In this Letter we develop theoretically, and demonstrate
experimentally, a new kind of general SOC in which the
orbital-angular-momentum (OAM) of atoms' center-of-mass couples to
their internal spin state, here referred to as
spin-orbital-angular-momentum coupling (SOAMC).  Amusingly, SOAMC is
closer to the original meaning of ``spin-orbit coupling" in atomic
physics where the OAM of an electron in an atom couples to the
electron spin. The SOAMC we report can be described by the
Hamiltonian,
\begin{eqnarray}
\hat{H_0}=\hat{h}_{\hat 1} +\hbar \delta \hat{F}_z+\hbar \Omega
\hat{F}_x +\hat{H}_{\rm
SOAMC}+\frac{\hbar^2}{2mr^2}\hat{F}_z^2,\label{eq:Hsoamc}
\end{eqnarray}
where $\hat{h}_{\hat 1}=-(\hbar^2/2m)\nabla^2(r,z) +L_z^2/2m r^2$,
$\nabla^2(r,z)=r^{-1}\partial_r(r\partial_r)+\partial^2_z,~L_z= -i
\hbar
\partial_{\phi}$ is the canonical angular momentum, $\hat{F}_i$ are the
spin-1 matrices, $\delta$ and $\Omega$ are the effective magnetic
fields along ${\mathbf e}_{z}$ and ${\mathbf e}_{x}$, and
$\hat{H}_{\rm SOAMC}=(\hbar/mr^2) L_z \hat{F}_z$ is the SOAMC term.
Here $\hat{H_0}$ arises from the laboratory Hamiltonian
$\hat{H}_{\rm lab}$ after a local spin rotation (see supplement),
where $\hat{H}_{\rm lab}=-(\hbar^2/2m)\nabla^2+ \vec{\Omega}_{\rm
eff}\cdot \vec{F}$, and $\vec{\Omega}_{\rm eff}$ is the local
light-induced effective magnetic field~\cite{Goldman2013} from
vector light shifts.

\begin{figure*}
    \includegraphics[width=7in]{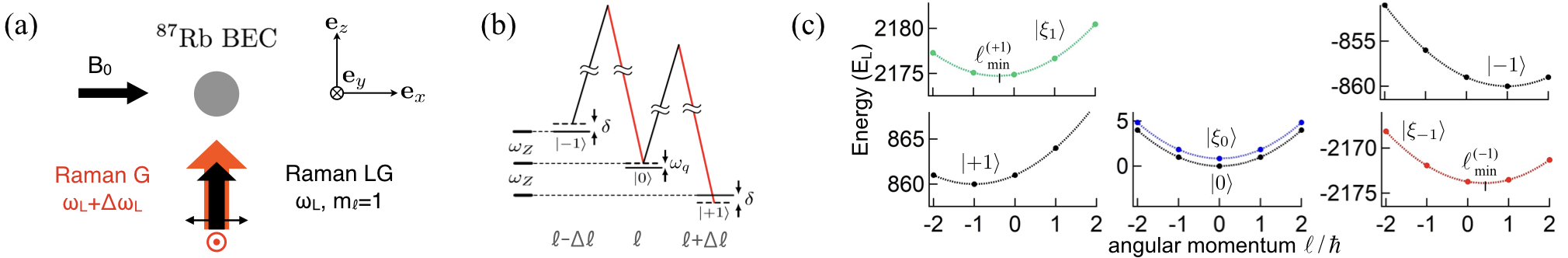}
    \caption{(a) Experimental setup (b) Level diagram showing the Raman beams transfer OAM= $\Delta \ell=\hbar$ between spin state $|m_F\rangle \leftrightarrow |m_F+1\rangle$(c) Energy dispersion $E(\ell)$ at $r=r_0=5~\mu$m without the quadratic Zeeman energy. The black symbols indicate dispersions of bare $|m_F\rangle$ without Raman coupling (see Eq.~\eqref{eq:Hsoamc} and supplement).Green, blue, and red symbols represent dispersions ofthe dressed states $|\xi_{n}\rangle$ with minima at $\ell_{\rm min}^{(n)}(\Omega(r_0),\delta)= r_0 A_{n}$, which can be tuned continuously and $n=\pm 1,0$. Energies are in unit of $E_{L}=\hbar^2/2mr_{0}^2=h\times 2.326~$Hz; $(\Omega(r_0),\delta)=2\pi\times(4.6,2.0)$~kHz.}
\end{figure*}

SOAMC systems~\cite{Juzeliunas2005} with $\hat{H_0}$ have azimuthal
gauge potentials $A {\mathbf e}_{\phi}$ owing to the coupling
between $\hat{F}_z$ and $L_z= -i \hbar
\partial_{\phi}$. The stationary Hamiltonians with $A {\mathbf e}_{\phi}$
are equivalent to the Hamiltonians in rotating frames, where $A
{\mathbf e}_{\phi}$ is an effective rotation. One can study the
properties at equilibrium with time-independent potentials, which is
impossible for systems under mechanical rotations with imperfect
cylindrical symmetries. This study differs from those where
metastable superflows were
investigated~\cite{Ramanathan2011,Wright2013,Beattie2013}. The
effective rotation with SOAMC can be used to measure superfluid
fractions ~\cite{Cooper2010} using the gauge-dependent spin
population imbalance of the dressed states, which vanishes in the
gauge of SLMC (see supplement~\footnote{See Supplementary Materials
for gauge dependence,high order Laguerre-Gaussian beams, magnetic
field control, BEC production,imaging calibration, and deloading,
which include
Refs.~\cite{LeBlanc2015,Moulder2013,Trypogeorgos2018,Lin09a,Reinaudi2007,lin11a}}.).
The gauge of SOAMC differs from that in SLMC, allowing one to probe
rotational properties under cylindrically symmetric configurations.

The engineered SOAMC also allows investigation of topological
excitations~\cite{Kawaguchi2012,Ueda2014} with cylindrical symmetry
in spinor BECs, e.g., coreless vortices~\cite{Leanhardt2003},
skyrmions~\cite{Choi2012a} and monopoles~\cite{Ray2014}. Such spin
textures can be created with SOAMC, but not with SLMC. The many-body
physics in SOAMC systems is rich and worth exploring in its own
right, in analogy to SLMC systems with spatially uniform Raman
coupling strength and Raman detuning. Examples that have been
discussed in SLMC include effective two-body
interactions~\cite{Williams12}, related physics with optical
lattices~\cite{Cole2012,Radic2012}, and ground state phases with
spin-dependent interactions in the small Raman coupling
regime~\cite{Goldman2013,Zhai2015}. The last is also considered in
Ref.~\cite{Qu2015,Sun2015,Chen2016,DeMarco2015,Hu2015} for SOAMC. In
this work, we consider $\hat{H}_{\rm SOAMC}$ where only the spin
component $\hat{F}_z$ is coupled, analogous to the $k_x \hat{F}_z$
in SLMC. With our technique using Laguerre-Gaussian (LG) Raman beams
to induce SOAMC, it is generally possible to create more complex
forms by versatile engineering of the Raman beams, such as those
with non-abelian gauge potentials~\cite{Dalibard11,Goldman2013}.

For dressed atoms under sufficiently large atom-light coupling
$\vec{\Omega}_{\rm eff}\cdot \vec{F}$, the eigenstates of the
overall Hamiltonian are well approximated as the local dressed spin
states, whose quantization axis is along $\vec{\Omega}_{\rm eff}$.
Thus one can employ a slowly varying position-dependent
$\vec{\Omega}_{\rm eff}$ to load the atoms into the dressed spin
state by adiabatic following. This is equivalent to manipulating
atoms using the spin rotation
method~\cite{Leanhardt2003,Choi2012a,Ray2014,Ollikainen2017} with
the Hamiltonian term $\vec{B}\cdot \vec{F}$ where $\vec{B}$ is a
``real" magnetic field~\cite{Isoshima2000}. An approach utilizing a
light-induced $\vec{\Omega}_{\rm eff}$ can thus allow versatile
design of $\vec{\Omega}_{\rm eff}$ to study topological excitations
in spinor BECs. The rich variety of order parameters in spinor BECs
can accommodate various types of topological
defects~\cite{Kawaguchi2012,Ueda2014}. Realizations of topological
excitations include those using spin rotation
methods~\cite{Leanhardt2003,Choi2012a,Ray2014,Choi2012,Ray2015,Hall2016,Ollikainen2017}
and Raman pulses of LG beams~\cite{Wright2009,Leslie2009}. The
topological excitations in the $|\langle \vec{F} \rangle|=1$
manifold have spin pointing along $\vec{\Omega}_{\rm eff}$, a
cylindrically symmetric configuration that cannot be created by
$\vec{\Omega}_{\rm eff}$ of SLMC.

We demonstrate light-induced SOAMC in atomic BECs by dressing the
atoms with a pair of Raman laser beams, one of which is an LG beam
carrying OAM. The beams couple atoms between different spin states
while transferring OAM from the light to the atoms' center-of-mass.
We adiabatically load a $^{87}$Rb BEC in $|F=1, m_F=0\rangle$ into
the $\langle \vec{F} \rangle=0$ (polar phase~\cite{Ho1998})
Raman-dressed state with light-induced $\vec{\Omega}_{\rm eff}$,
where coreless vortices~\cite{Kawaguchi2012,Ueda2014} are created.
Each decomposed bare spin state $|m_F\rangle$ of the dressed state
has its (orbital) angular momentum correlated with $m_F$, indicating
SOAMC. Similar spin textures were reported in
Refs.~\cite{Leanhardt2003,Choi2012a} with the spin rotation method.
Here we characterize the atoms' temporal evolutions with the Raman
dressing field and after turning it off. We observe that the
middle-energy dressed state is stable in the Raman field for
$\approx~0.1$ s at Raman resonance, and that its lifetime is
prolonged for larger Raman detunings. When the Raman field is turned
off, we observe the vortices in the bare spin $|m_F=\pm1 \rangle$
components split into half-vortices~\cite{Seo2016}, after which $|1
\rangle$ and $|-1\rangle$ components spatially separate.

Consider the laboratory Hamiltonian $\hat{H}_{\rm lab}$ with
$\vec{\Omega}_{\rm eff}=\Omega(r)\cos \phi{\mathbf
e}_{x}-\Omega(r)\sin \phi{\mathbf e}_{y}+\delta{\mathbf e}_{z}$ in
our setup. This Hamiltonian is transformed to $\hat{H_0}$ under a
local spin rotation to remove the $\phi$-dependence of
$\vec{\Omega}_{\rm eff}$, giving rise to $\hat{H}_{\rm SOAMC}$ and
$(\hbar^2/2mr^2)\hat{F}_z^2$. Consider first the atoms with no
motional degree of freedom, where the dressed
eigenstates~\cite{Dalibard11,Goldman2013,Spielman2013a} are exactly
given by diagonalizing $\vec{\Omega}_{\rm eff}\cdot \vec{F}$ and are
the dressed spin states with the quantization axis along
$\vec{\Omega}_{\rm eff}(\vec{r},t)$. We then include the motional
kinetic energy $-(\hbar^2 /2m) \nabla^2$, and consider a general
atomic state $\langle
\vec{r}|\Psi(t)\rangle=\psi(\vec{r},t)|\xi(\vec{r},t)\rangle$, where
$\psi$ is the external part and $|\xi\rangle$ is the (normalized)
spin part of the wave function. If the ``adiabatic condition" is
fulfilled, the atoms can be initially prepared in a particular
branch of dressed states and remain in the same state as time
evolves. The adiabatic condition is satisfied by sufficiently slowly
varying external parameters and small spatial gradient in both
$\psi$ and $|\xi\rangle$. More explicitly, with small spatial
gradient energies compared to the gap $|\vec{\Omega}_{\rm
eff}|=\Omega_{\rm eff}$ between dressed states, it is valid to take
$\hat{H}_{\rm SOAMC}$, $(\hbar^2/2mr^2)\hat{F}_z^2$ and terms with
$\partial_r$ as perturbations. Thus $\vec{\Omega}_{\rm eff}\cdot
\vec{F}$ is the dominating term in the Hamiltonian.

Given the adiabatic condition for atoms in the $n-\rm{th}$ dressed
state, the atomic state can be expressed as $\langle
\vec{r}|\Psi(t)\rangle=\psi_n(\vec{r},t)|\xi_n(\vec{r},t)\rangle$,
where $|\xi_n (\vec{r},t)\rangle$ are the local dressed spin states
and are normalized spinor eigenstates of $\vec{\Omega}_{\rm
eff}\cdot \vec{F}$ in states $n=0,\pm 1$; $\psi_n$ are the external
wave functions. We then consider the projected
Hamiltonian~\cite{Goldman2013} $H_{\rm eff}^{(n)}= \langle
\xi_n|\hat{H}_{\rm lab}|\xi_n\rangle +V(r)$ which governs the
dynamic motion $\psi_n(\vec{r},t)$ and a gauge potential
$\vec{A}_{n}=A_{n}(r,t){\mathbf e}_{\phi}= (i\hbar/r)\langle
\xi_n|\partial_{\phi} \xi_n\rangle {\mathbf e}_{\phi}$ appears,
\begin{eqnarray}
H_{\rm eff}^{(n)}=\frac{-\hbar^2}{2m}\nabla^2(r,z)
+\frac{\left(L_z-r A_{n}\right)^2}{2m r^2} + V(r)+\varepsilon_n+W_n,
\label{eq:projectedH}
\end{eqnarray}
 $V(r)$ is the spin-independent trap, $\varepsilon_n=n
\Omega_{\rm eff}$ is the eigenenergy of $\vec{\Omega}_{\rm eff}\cdot
\vec{F}$ and $W_n$ is the geometric scalar potential $\approx
\hbar^2/2mr^2$. At sufficiently large $r$ under the adiabatic
condition, terms with $\partial_r$ are negligible. We then consider
the effective energy dispersion at fixed $r$ vs. $\ell$, which is
the eigenvalue of $L_z$ quantized in units of $\hbar$. The
dispersion adds up $(\ell-r A_{n})^2/2mr^2$, $\varepsilon_n$ and
$W_n$ (Fig.1c). We label the lowest, middle and highest energy
dressed states as $|\xi_{-1}\rangle, |\xi_0\rangle$ and
$|\xi_1\rangle$, respectively. The dressed atoms in $|\xi_n\rangle$
have kinematic angular momentum $\ell-\ell_{\rm min}^{(n)}$, where
$\ell_{\rm min}^{(n)}=r A_n$.

\begin{figure}
    \centering
    \includegraphics[width=3.4 in]{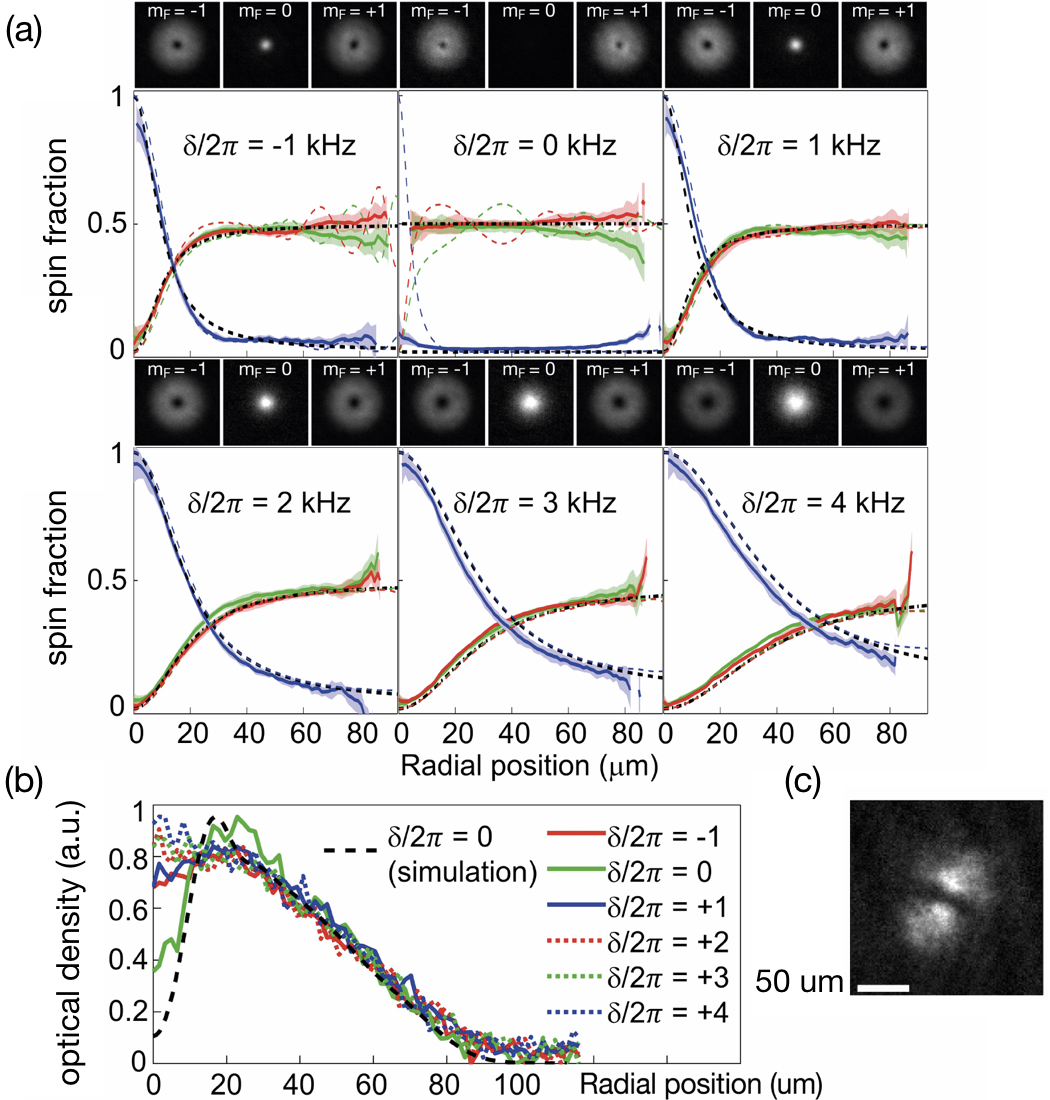}
    \caption{Demonstration of SOAMC in the vertical images of the dressed state $|\xi_0\rangle$ after 24 ms TOF with $t_h=1$~ms.
    (a) Images projected onto bare spin
    components $|m_F\rangle$ for $-1$~kHz$<\delta/2\pi<4$~kHz and the radial cross
    sections of spin textures versus theory. The image scale is $240~\mu$m$\times 240~\mu$m.
    The blue, red and green curves are data for $|0\rangle,|1\rangle$ and $|-1\rangle$ components, respectively,
     with the shaded region indicating the uncertainty. The black dashed (dash-dotted)
    curves indicate predictions from Eq.~\eqref{eq:spinor} magnified by 9.1 in the radial position~\cite{Castin1996} for
    $|0\rangle (|\pm 1\rangle)$. The colored dashed curves indicate TOF simulations from 3D TDGPE. (b) Radial cross sections
     of the total optical density of all $|m_F\rangle$ for all $\delta$. Colored solid and dotted (black dashed) curves denote
     the data (TOF simulation at $\delta=0$). (c) Interference between $|-1\rangle$ and $|+1\rangle$ components at $\delta=0$.
    }

\end{figure}

We start with a $^{87}$Rb BEC in $|F=1, m_F=-1\rangle$ state in a
crossed optical dipole trap with $N\approx~4 \times 10^5$ atoms. The
condensate is approximately spherical-symmetric with a Thomas-Fermi
(TF) radius $R_{\rm TF}\approx 10~\mu$m. A bias magnetic field $B_0$
along ${\mathbf e}_{x}$ gives a linear Zeeman shift
$\omega_Z/2\pi=0.57$~MHz and a quadratic Zeeman shift $\hbar
\omega_q \hat{F}_z^2$ with $\omega_q/2\pi= 50$ Hz. Two $\lambda=
790$ nm Raman laser beams co-propagate along ${\mathbf e}_{z}$. One
is a Gaussian beam (G), and the other is an LG beam with phase
winding $m_{\ell}=1$. The Raman beams transfer
OAM=$\triangle\ell=\hbar$ when coupling the atoms from $|m_F\rangle$
to $|m_F+1\rangle$ (Fig.1a 1b). The Raman laser frequencies differ
by $\Delta \omega_L$, and the Raman detuning is $\delta=\Delta
\omega_L- \omega_Z$. We transfer the BEC to $|m_F=0 \rangle$ and
then load the atoms into the $|\xi_0\rangle$ dressed state with
$\Omega(r,t)$ and $\delta(t)$. The final value of Raman coupling
strength is $\Omega(r)=\Omega_M \sqrt{e}(r/r_M) e^{-r^2/2r_M^2}$
where $\Omega_M/2\pi= 10$~kHz; $r_{M}=17~\mu$m is the radius at peak
intensity. The resulting $\vec{\Omega}_{\rm eff}$ has the polar
angle $\beta(r)=\tan^{-1} [ \Omega(r)/\delta]$.

We load the atoms into the dressed state by turning on $\Omega(r)$
in 15~ms at detuning $\delta/2\pi=5$~kHz, followed by ramping
$\delta/2\pi$ to between 4 kHz and -1 kHz. The resulting spinor
state is the local dressed state $|\xi_0\rangle$ for $r
> r_c$ where the adiabatic condition is fulfilled. Here $r_c$ is the
adiabatic radius determined by the loading speed $\dot{\delta}$ and
spatial gradient energies with respect to $\Omega_{\rm eff}(r)$. The
quadratic Zeeman shift is smaller than $\Omega_{\rm eff}(r)$ or
spatial gradient energies, thus its effect is negligible. We perform
3D time-dependent-Gross-Pitaevskii equation (TDGPE) simulations
including the kinetic energies, quadratic Zeeman energy, mean field
interaction parameters $c_0=4\pi \hbar^2(a_0+2a_2)/3m$ and $c_2=4\pi
\hbar^2(a_2-a_0)/3m<0$, where $a_f$ is the s-wave scattering length
in the total spin $f$ channel~\cite{Ho1998}. This gives an $r_c
\approx 1.4~\mu$m at resonance $\delta=0$.

To a good approximation, for $r> r_c$ the condensate can be loaded
into the local dressed state~\cite{Ho1998},
\begin{eqnarray}
|\xi_{0}\rangle &= \left[ -e^{i \phi}\sin \beta(r)/\sqrt{2}, \cos
\beta(r), e^{-i \phi}\sin \beta(r)/\sqrt{2} \right]^{\rm
T},\label{eq:spinor}
\end{eqnarray}
where the local quantization axis is along $\vec{\Omega}_{\rm eff}$
with $A_{0}=0$ during loading. The wave functions of bare spin
$|m_F\rangle$ components possess spin-dependent angular momentum
$\ell+m_F\hbar$, showing the SOAMC; here $\ell=0$. For atoms in
$|\xi_{\pm 1}\rangle$, $A_{\pm 1}= \mp(\hbar/r)\cos \beta(r)$ (see
Fig.1c) are the azimuthal gauge potentials.

Following the preparation, we probe the dressed state after a hold
time $t_h$ by switching off the Raman beams and dipole trap
simultaneously, and then adiabatically rotating the bias field to
the direction along the imaging beam. After a 24 ms time-of-flight
(TOF), we take images along the $z$ direction of each spin $m_F$
state respectively, by using microwave spectroscopy for
$m_F$-selected imaging. A single $m_F$ state is imaged in an
experimental realization. With all spin states expanding together
during TOF, in the $|c_2|\approx 0$ limit each $|m_F\rangle$ after
TOF approximately experiences a self-similar dilation of the in-situ
profile by the same factor~\cite{Castin1996}, verified by the TDGPE
simulations; the in-situ and TOF profiles largely agree at $r
\gtrsim 2~\mu$m. The only exception is for $\delta \lesssim 0$,
where the $|\pm 1\rangle$ components show oscillations of spin
imbalance versus $r$ after TOF (Fig.~2a). The images for different
detuning $\delta$ with a short $t_h=1$~ms are shown in Fig.~2a,
indicating $|0\rangle$ carries zero angular momentum and $|\pm
1\rangle$ carry the same magnitude of angular momentum from their
same hole sizes. To prove $|+1\rangle,|-1\rangle$ carry $\pm\hbar$
(or $\mp\hbar$), we take the interference between the two components
at $\delta=0$ where the nodal-line shows the $4\pi$ relative phase
winding (Fig.~2c). The spin-dependent angular momentum demonstrates
SOAMC. Fig.~2a shows radial cross sections of the spin texture
$D_{m_F}/(D_1+D_0+D_{-1})$ after averaging over the azimuthal angles
compared to Eq.~\eqref{eq:spinor} and TOF simulations, where
$D_{m_F}$ is the optical density of $|m_F\rangle$. At $\delta=0$ the
population in $|0\rangle$ reaches the minimum; at the TOF position
$r_{\rm TOF}\gtrsim 9~\mu$m corresponding to the in-situ $r>1~\mu$m,
the data does not show the predicted oscillations of spin imbalance,
likely due to small dissipations in the experiment. The profile at
$\delta/2\pi=1$~kHz is similar to that at $\delta/2\pi=-1$~kHz for
$r > r_c$ as expected. Our experimental results agree with the
prediction for -1~kHz$<\delta/2\pi<4$~kHz.

\begin{figure}
    \centering
    \includegraphics[width=3.4 in]{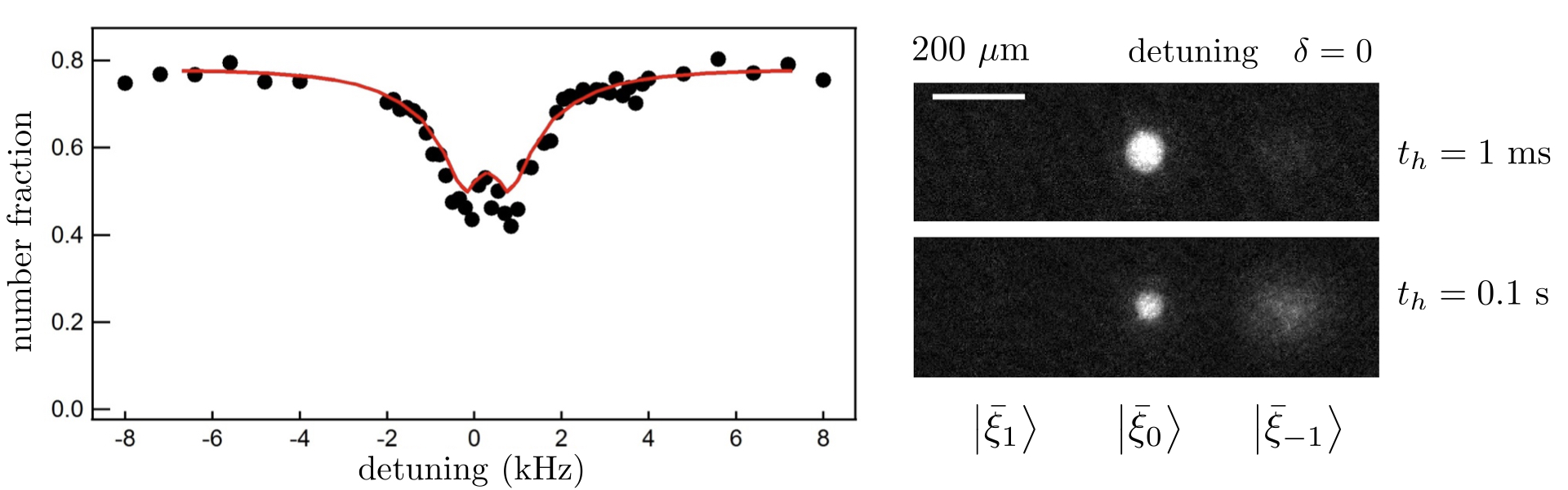}
    \caption{Number fraction in the dressed state $|\xi_0\rangle$ after a $t_h=0.1$~s hold
    time with Raman fields on versus detuning $\delta$, compared with the calculated
    rate (red curve). The inset shows the side images at $\delta=0$ for $t_h=1$~ms and 0.1~s. }
\end{figure}

We theoretically discuss the stability of the dressed atoms with the
Raman field for $t_h>0$. Without the interaction, the total
Hamiltonian after the local spin rotation is $\hat{H_1}=\hat{H_0}+
V(r)$. By neglecting terms with $\partial_r$, the eigenstates at
fixed $r$ with the good quantum number $\ell$ are the ``modified
local dressed states" $|\bar{\xi}_n(\ell,r)\rangle$, which deviate
from $|\xi_n\rangle$ due to $L_z^2/2mr^2$ and last two terms in
Eq.~(1). While at $\ell=0$, $|\bar{\xi}_n(\ell,r)\rangle$
approximates $|\xi_n\rangle$ for $r>r_c$. With our $\dot{\delta}$,
for $r\gtrsim 1.4~\mu$m the atoms are adiabatically prepared in
$|\xi_0\rangle \approx |\bar{\xi}_0\rangle$ at $\ell=0$. From TDGPE
simulations we find $|\bar{\xi}_0\rangle$ is coupled to
$|\bar{\xi}_{-1}\rangle$ after a hold time with $\ell=0$ unchanged
(see supplement). Beyond the mean field description, we consider
interactions in the second quantization form, $\Hat{H}_{\rm
int}^{c_0}$, where the dominating spin-independent $c_0$ term can
make the dressed atoms decay to the lowest energy state
$|\bar{\xi}_{-1}\rangle$ in the presence of general
SOC~\cite{Spielman09,Zhang13}. In contrast, the $c_0$ term cannot
couple between bare spin states.

We study the stability of the spinor state initially loaded into
$|\xi_0\rangle \approx |\bar{\xi}_0\rangle$ with the Raman field. To
investigate time-evolving distributions in dressed states, we
perform deloading after a variable hold time $t_h$; for $r>r_c$ we
map the dressed states ${|\bar{\xi}_{-1}\rangle,
|\bar{\xi}_{0}\rangle, |\bar{\xi}_{1}\rangle}$ back to the bare spin
states ${|-1\rangle, |0\rangle, |+1\rangle}$, respectively. We take
side images after 14 ms-TOF with Stern-Gerlach gradient versus
$\delta$ with $t_h=1$~ms and 0.1~s. We display the fraction of the
atom number in $|\bar{\xi}_{0}\rangle$ over the total number in
$|\bar{\xi}_{0}\rangle$ and $|\bar{\xi}_{-1}\rangle$ at $t_h=0.1$~s
(Fig.~3). We compare the data with the simulation of decay (see
supplement) from $|\bar{\xi}_{0}\rangle$ to $|\bar{\xi}_{-1}\rangle$
due to the collisions $\Hat{H}_{\rm int}^{c_0}$, and the two agree
(Fig.~3).

Finally, we investigate evolutions of the initial dressed state
after turning off the Raman fields. TOF vertical images are taken
after leaving the sample in the dipole trap with a hold time $t_{\rm
off}$. We observe the $|m_F=\pm 1\rangle$ vortices repel each other,
and the domains in magnetization images show spatial separation. At
$t_{\rm off}\geq 50$~ms the images vary in different experimental
shots under identical conditions, likely due to variations of vortex
centering in the shots and dynamical instabilities. Similar dynamics
has also been studied in Ref.~\cite{Choi2012a}.

\begin{figure}
    \centering
    \includegraphics[width=3.4 in]{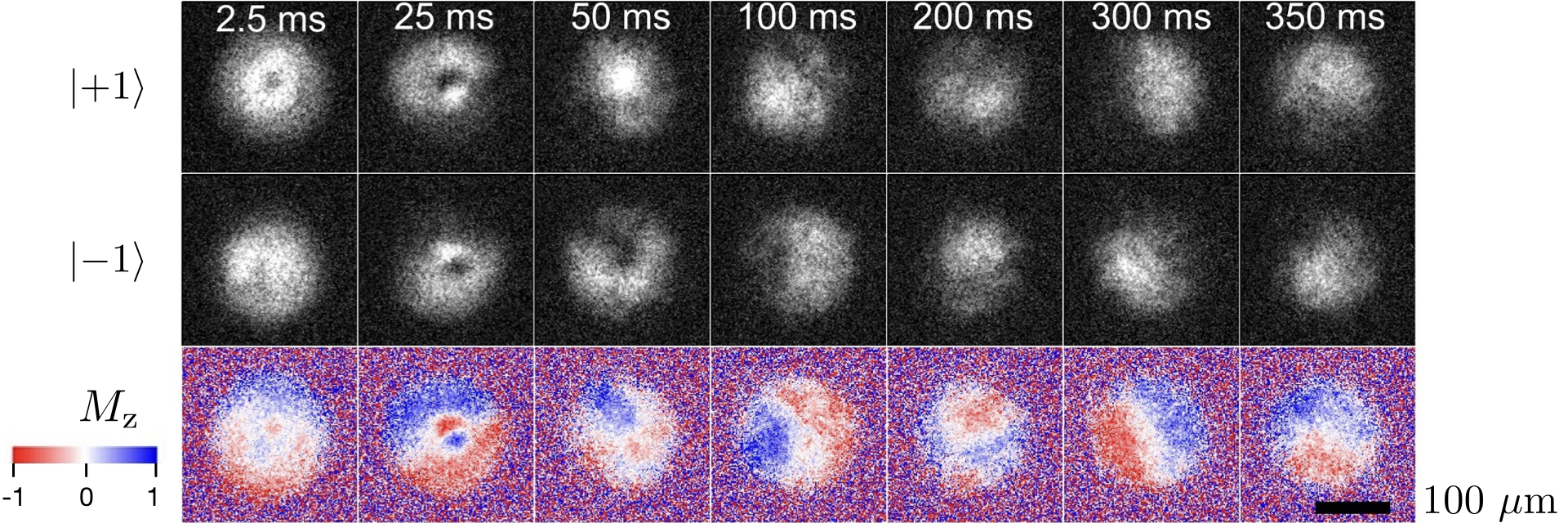}

        \caption
    {Time evolution of the initially prepared dressed state at $\delta=0$
    after a sudden turn-off of the Raman fields and holding $t_{\rm
    off}$, $2.5$~ms $\leq t_{\rm off}\leq$ 350~ms.
    The images in the top (middle) row are projected to $|m_F=\pm 1\rangle$;
    the $|m_F=0\rangle$ components are not discernible. The images
    are from single experiments and are selected to best illustrate
    the evolution. The bottom row
    shows magnetization images $M_z=(D_{1}-D_{-1})/(D_{1}+D_0+D_{-1})$.
    }

\end{figure}

In conclusion, we demonstrate SOAMC by creations of coreless
vortices in the $F=1$ polar phase BEC by loading the atoms into
Raman dressed states. We also study their stability with and without
the Raman fields. Going beyond our initial demonstration, the
transverse components of light-induced $\vec{\Omega}_{\rm eff}$ can
be engineered via intensity and phase patterns of the Raman beams
using spatial-light-modulators, and the axial component can be
manipulated via vector light shifts from another laser. Designing of
$\vec{\Omega}_{\rm eff}$ enables smaller spatial scales and faster
time scales than using a real magnetic field $\vec{B}$, and opens
more possibilities for creating topological structures.
Manipulations with $\vec{\Omega}_{\rm eff}$, instead of $\vec{B}$,
allow independent control of $\vec{B}$. With nonzero $\nabla\cdot
\vec{\Omega}_{\rm eff}$, one can create a synthetic ``antimonopole"
with opposite charge to that generated by $\vec{B}$ in
Refs.~\cite{Ray2014,Ray2015}. The interaction of a
monopole-antimonopole pair or a vortex pair with controlled pair
sizes can be studied. With non-collinear vortices, their collisions,
cutting and reconnections can be studied. For non-abelian vortices
in the $F=2$ manifold, the production of rung vortex
~\cite{Kawaguchi2012,Ueda2014} can be tested. Further, generating
high-order LG Raman beams may reach larger $E_L\propto {\Delta
\ell}^2$ and potentially access the small Raman coupling regime with
multiple minima in the energy dispersion and the predicted miscible
annular stripe phases~\cite{Qu2015,Sun2015,Chen2016} (see
supplement).

\begin{acknowledgments}
The authors thank M.~-S. Chang, C. Chin, I.~B. Spielman, B. Xiong
and J.~T. Hougen for useful discussions, and W.~D. Phillips for
critical readings of the manuscript. We also thank H.~-J. Wei and
C.~-Y. Yu for their contributions to build the experiment. Y.~-J.~L.
was supported by MOST, Career Development Awards in Academia Sinica
and NCTS ECP1. S.~-K.~Y. was supported by MOST. Y. K. was supported
by JSPS KAKENHI Grant Numbers JP15K17726 and JP16H00989.
\end{acknowledgments}

%%% combine supplementary: https://tex.stackexchange.com/questions/168169/options-for-supplementary-materials-in-preprint-version-revtex-arxiv 

\widetext
\clearpage
\begin{center}
\textbf{\large Supplemental Materials\\ Spin-orbital-angular-momentum coupled Bose-Einstein condensates}
\end{center}
%%%%%%%%%% Merge with supplemental materials %%%%%%%%%%
%%%%%%%%%% Prefix a "S" to all equations, figures, tables and reset the counter %%%%%%%%%%
\setcounter{equation}{0}
\setcounter{figure}{0}
\setcounter{table}{0}
\setcounter{page}{1}
\makeatletter
\renewcommand{\theequation}{S\arabic{equation}}
\renewcommand{\thefigure}{S\arabic{figure}}
\renewcommand{\bibnumfmt}[1]{[S#1]}
\renewcommand{\citenumfont}[1]{S#1}
%%%%%%%%%% Prefix a "S" to all equations, figures, tables and reset the counter %%%%%%%%%%

\section{Formalism of dressed states}
\subsection{Hamiltonian with SOAMC}
With the bias field along ${\mathbf e}_{x}$ and taking the
conventional quantization axis along ${\mathbf e}_{z}$, we perform a
global spin rotation, $\Hat{F}_x\rightarrow
\Hat{F}_z,\Hat{F}_y\rightarrow \Hat{F}_x,\Hat{F}_z\rightarrow
\Hat{F}_y$. We then make the rotating wave approximation, and the
Hamiltonian in the bare spin basis
${|+1\rangle,|0\rangle,|-1\rangle}$ in the frame rotating at $\Delta
\omega_L$ is
\begin{align}\label{eq:Hlab}
\hat{H}_{\rm lab}=\left[\frac{-\hbar^2}{2m}\frac{\partial}{r\partial
r}(r \frac{\partial}{\partial
r})-\frac{\hbar^2}{2m}\frac{\partial^2}{\partial z^2}
+\frac{L_z^2}{2m r^2}\right] \otimes {\hat 1}+\vec{\Omega}_{\rm
eff}\cdot \vec{F} \nonumber\\
=\left[\frac{-\hbar^2}{2m}\frac{\partial}{r\partial r}(r
\frac{\partial}{\partial
r})-\frac{\hbar^2}{2m}\frac{\partial^2}{\partial z^2}
+\frac{L_z^2}{2m r^2}\right]
\otimes {\hat 1}+ \hbar \delta \hat{F}_z \nonumber\\
+\hbar\Omega(r)\cos{\phi}\Hat{F}_x-\hbar\Omega(r)\sin{\phi}\Hat{F}_y
\end{align}
in the $(r,\phi,z)$ coordinate. Here, $\vec{\Omega}_{\rm
eff}=\Omega(r)\cos \phi{\mathbf e}_{x}-\Omega(r)\sin \phi{\mathbf
e}_{y}+\delta{\mathbf e}_{z}$ given the OAM transfer
$\Delta\ell=\hbar$. We perform a local spin rotation about ${\mathbf
e}_{z}$ by the azimuthal angle $-\phi$ to remove the
$\phi-$dependence of $\vec{\Omega}_{\rm eff}$, making
$\vec{\Omega}_{\rm eff}\cdot \vec{F}$ transformed to $\hbar \delta
\Hat{F}_z+\hbar \Omega \Hat{F}_x$, and thus
\begin{align}\label{eq:H0}
\hat{H}_0=\left[\frac{-\hbar^2}{2m}\frac{\partial}{r\partial r}(r
\frac{\partial}{\partial
r})-\frac{\hbar^2}{2m}\frac{\partial^2}{\partial z^2}
+\frac{L_z^2}{2m r^2}\right]
\otimes {\hat 1} \nonumber\\
+\hbar \delta \Hat{F}_z+\hbar \Omega \Hat{F}_x+\hat{H}_{\rm
SOAMC}+\frac{\hbar^2}{2mr^2}\Hat{F}_z^2,
\end{align}
where $\hat{H}_{\rm SOAMC}=(\hbar/mr^2)L_z \Hat{F}_z$. This can be
expressed as
\begin{align}
\hat{H}_0=\hat{h}_0 \otimes {\hat 1}+\hbar \delta \Hat{F}_z+\hbar
\Omega \Hat{F}_x+\left(
\begin{array}{ccc}
    (L_z+\hbar)^2/(2mr^2)     & 0  & 0\\
    0                & L_z^2/(2mr^2)        &0 \\
    0                & 0            & (L_z-\hbar)^2/(2mr^2)
\end{array}\right),
\end{align}
where
$\hat{h}_0=-(\hbar^2/2m)\left[r^{-1}\partial_r(r\partial_r)+\partial^2_z\right]+V(r)$,
and the $3\times 3$ matrix indicates the spin-$m_F$-dependent
azimuthal kinetic energy $(L_z+m_F \hbar)^2/(2mr^2)$ for
$m_F=\pm1,0$. This shows the energy dispersion for bares spin state
$|m_F\rangle$ in Fig.~1c is $(L_z+m_F \hbar)^2/(2mr^2)+m_F \hbar
\delta$.

Finally with a global spin rotation, $\Hat{F}_z\rightarrow
\Hat{F}_x,\Hat{F}_x\rightarrow \Hat{F}_y,\Hat{F}_y\rightarrow
\Hat{F}_z$, it gives
\begin{align}
\hat{H}=\left[\frac{-\hbar^2}{2m}\frac{\partial}{r\partial r}(r
\frac{\partial}{\partial
r})-\frac{\hbar^2}{2m}\frac{\partial^2}{\partial z^2}
+\frac{L_z^2}{2m r^2} \right]
\otimes {\hat 1} \nonumber\\
+\hbar \delta \Hat{F}_x+\hbar \Omega \Hat{F}_y+ \frac{\hbar }{mr^2}
L_z \Hat{F}_x+\frac{\hbar^2}{2mr^2}\Hat{F}_x^2,
\end{align}
back to the quantization axis along ${\mathbf e}_{x}$.

\subsection{Gauge potentials}
For atoms in bare spin state $|m_{F}=n\rangle$ and are adiabatically
loaded to the dressed state $|\xi_n(\vec{r},t)\rangle$, this can be
described with an Euler rotation~\cite{Ho1998s} with the Euler angles
$(\alpha,\beta,\gamma)$,
\begin{align}
|\xi_n(\vec{r},t)\rangle=\mathcal{U}(\alpha,\beta,\gamma)|m_{F}=n\rangle,\label{eq:Euler}
\end{align}
where $n=\pm 1(0)$ is for the ferromagnetic $|\langle \vec{F}
\rangle|=1$ (polar $\langle \vec{F} \rangle=0$) state, $\alpha$ and
$\beta$ are given by the azimuthal and polar angle of
$\vec{\Omega}_{\rm eff}(\vec{r},t)$, respectively. For the $n=\pm 1$
ferromagnetic state, $\gamma$ is equivalent to the gauge choice
while for the $n=0$ polar state $\gamma$ does not appear. This leads
to
\begin{align}
|\xi_{\pm1}(r,t)\rangle= e^{\mp i \gamma}\left(
\begin{array}{c}
{e^{-i \alpha}} \frac{1 \pm \cos \beta}{2}\\
\frac{\pm 1}{\sqrt{2}} \sin \beta \\
{e^{i \alpha}} \frac{1 \mp \cos \beta}{2}
\end{array}\right),
|\xi_{0}(r,t)\rangle= \left(
\begin{array}{c}
{-e^{-i \alpha}} \frac{\sin \beta}{\sqrt{2}}\\
\cos \beta \\
{e^{i \alpha}} \frac{\sin \beta}{\sqrt{2}}
\end{array}\right).
\label{eq:spin1}
\end{align}
Two conventional choices of $\gamma$ for the ferromagnetic state are
$\gamma=0$ and $\gamma= \mp \alpha$, where both $\gamma$ and
$\alpha$ are time-independent (see next paragraph). With these
choices, the dynamical phase appears in the phase of the external
part of wave function $\psi_n(r,t)$, not in $|\xi_n(r,t)\rangle$.

In our $\vec{\Omega}_{\rm eff}(\vec{r},t)$ from the Gaussian and LG
Raman beams, $\alpha= -(\Delta\ell /\hbar) \phi$ is an integer
multiple of $\phi$ and time independent, consequently in
Eq.~\eqref{eq:spin1} the phase winding number (the integer given by
the phase gradient along ${\mathbf e}_{\phi}$) of each $|m_F\rangle$
component is stationary. Besides, since $\left[ H_{\rm
eff}^{(n)},L_z \right]=0$, the phase gradient of $\psi_n(\vec{r},t)$
along ${\mathbf e}_{\phi}$ remains zero and this gradient is
developed with time (initially zero) only along ${\mathbf
e}_{r}$~\cite{Kawaguchi2012s}.

For our experiment where $|\Psi\rangle$ was initially polarized in
$|m_F=0\rangle$ with zero angular momentum, $\gamma$ doesn't appear
in $|\xi_0\rangle$ and thus the gauge potential for $|\xi_0\rangle$
is $A_0=0$ without additional phase term of gauge transformation.
This corresponds to the conventional gauge choice of $\gamma=0$ for
$n=\pm 1$, leading to $A_{\pm 1}= \mp(\hbar/r)\cos \beta(r)$ for
$|\xi_{\pm 1}\rangle$ (see the later Eq.~\eqref{eq:gauge}).

\subsection{3D TDGPE simulations for spin textures}
We numerically simulate the dynamics by solving the three-component
3D time-dependent-Gross-Pitaevskii equation (TDGPE). We use the
Crank-Nicolson method and calculate in the system size of $(256)^3$
grid points with grid size $0.22~\mu$m. During TOF, we solve the
full 3D TDGPE for up to $\le 6$~ms at which the interatomic
interaction energy becomes less than 5 percent of the total energy.
The further evolution is calculated by neglecting the interaction
term. The results for the polar dressed state with a short hold time
$t_h=1$~ms are shown in Fig.~S1; our corresponding data is in
Fig.~2.

% 3.4 in for 2-column
% 4.2 in for preprint
\begin{figure}
    \centering
    \includegraphics[width=3.4 in]{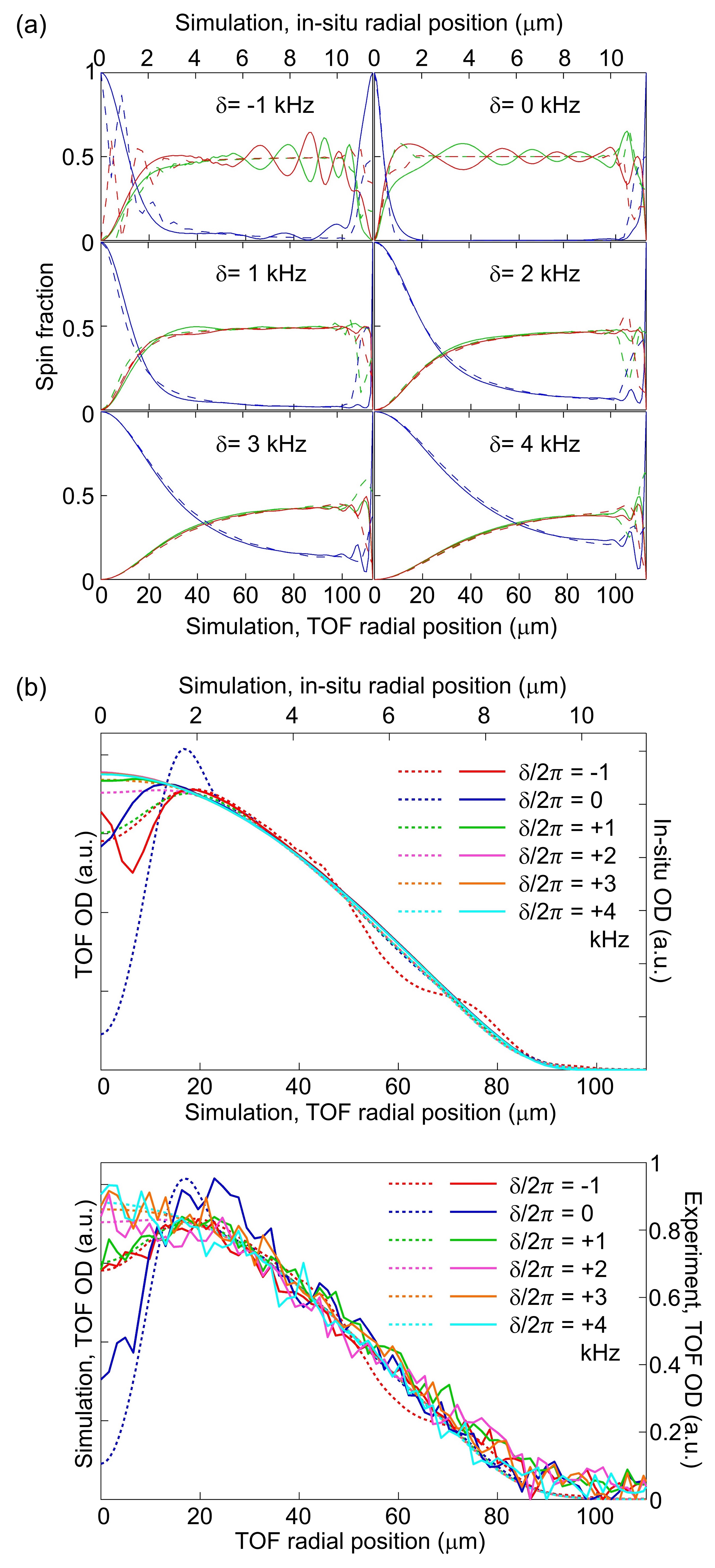}

        \caption
    {3D TDGPE simulation results for radial cross sections of polar dressed state
     at various detuning $\delta$ with a short hold time $t_h=1$~ms.
    (a) Spin texture $D_{m_F}/(D_1+D_0+D_{-1})$ for in-situ (dashed curves)
    and for after 24 ms TOF (solid curves). Blue, red and green
    curves denote $|0\rangle,|1\rangle,|-1\rangle$, respectively.
    Except for $\delta\leq 0$, the profiles at in-situ and after TOF agree well, i.e., it is a
    dilation after TOF for each $|m_F\rangle$.
    (b) Top panel: total optical density $(D_1+D_0+D_{-1})$ for
    simulated in-situ (solid curves) and TOF profiles (dashed
    curves). Bottom panel: total optical density for
    experimental (solid curves) and simulated TOF profiles (dashed
    curves).
    }

\end{figure}

\subsection{Dressed eigenstates and validity of the adiabatic condition}
The Hamiltonian $\hat{H}_{0}$ is dominated by the atom-light
coupling $\hbar \delta \Hat{F}_z+\hbar \Omega \Hat{F}_x$ at large
$r$ with a sufficiently large gap $\Omega_{\rm
eff}(r)=\sqrt{\Omega(r)^2+\delta^2}$. Thus the $\hat{H}_{\rm SOAMC}$
and $(\hbar^2/2mr^2)\Hat{F}_z^2$ originating from the gradient
energy $\hat{K}\equiv -(\hbar^2 /2m) \nabla^2 \otimes {\hat 1}$
after the local spin rotation can be treated as perturbations.

We consider the gradient energy being projected onto the basis of
local dressed states $|\xi_n\rangle$, where the off-diagonal term
$H_{n^{'}n}$ indicates coupling between dressed state $n$ and
$n^{'}$. We will prove the validity of local dressed states
$|\xi_n(\vec{r},t)\rangle$ as the approximated eigenstates, and of
the adiabatic condition, i.e., coupling between dressed states are
negligible.

Taking the Hamiltonian in Eq.~\eqref{eq:Hlab} and transform it to
that in the basis of local dressed state $|\xi_n(\vec{r},t)\rangle$,
the transformed Hamiltonian has gauge potential ${\mathbf A}=i \hbar
\mathcal{U}^\dag \nabla \mathcal{U}$~\cite{Spielman2013as} with
${\mathbf A}_{n^{'}n}=i\hbar\langle \xi_{n^{'}}|\nabla
\xi_n\rangle$,
\begin{align}\label{eq:gauge}
{\mathbf A}&=-\frac{\Delta\ell}{r}\cos{\beta(r)}\Hat{F}_z{\mathbf
e}_{\phi} +\frac{\Delta\ell}{r}\sin{\beta(r)}\Hat{F}_x{\mathbf
e}_{\phi}
+\hbar \partial_{r}\beta(r)\Hat{F}_y{\mathbf e}_{r},\nonumber\\
\vec{A}_{n}&={\mathbf A}_{nn} = -\frac{\Delta\ell}{r}\cos{\beta(r)}n
{\mathbf e}_{\phi},
\end{align}
\begin{align}\label{eq:H_offdiagonal0}
H_{n^{'}n} &=\frac{-\hbar}{2m}(k \cdot {{\mathbf
A}_{n^{'}n}}+{{\mathbf A}_{n^{'}n}} \cdot k)+\frac{ {\mathbf
A}_{n^{'}n} \cdot {\mathbf A}_{nn}+{\mathbf
A}_{n^{'}n^{'}} \cdot {\mathbf A}_{n^{'}n} }{2m},\\
k&=\frac{\nabla}{i}= \frac{1}{i}({\mathbf e}_{r}
\partial_r+{\mathbf e}_{\phi}\frac{\partial_{\phi}}{r})={\mathbf e}_{r} k_r+{\mathbf e}_{\phi}\frac{L_z}{\hbar r}\nonumber.
\end{align}
${\mathbf A}$ contains off-diagonal terms, and the diagonal term
$\vec{A}_{n}$ is the gauge potential for $|\xi_n\rangle$.
$\vec{A}_{n}$ results from the spatially-dependent $|\xi_n\rangle$;
it is contributed only from the phase gradient of the Raman
coupling's off-diagonal term $\Omega(r)e^{i\phi}$ in the spin matrix
in Eq.~\eqref{eq:Hlab}, and none from the amplitude gradient of
$\Omega(r)$. The phase of $\Omega(r)e^{i\phi}$ corresponds to
$\alpha=-\phi$ in Eq.~\eqref{eq:spin1} and the relative phase
between $\langle m_F-1|\xi_n\rangle$ to $\langle m_F|\xi_n\rangle$.
Therefore this relative phase gradient fixes the direction of
$\vec{A}_{n}$ long ${\mathbf e}_{\phi}$. With this, in Eq.~(2) the
scalar potential is $\varepsilon_n-i \hbar \langle \xi_n|\partial_t
\xi_n\rangle=\varepsilon_n$ given by a general $\beta(r,t)$ and
time-independent $\alpha=-\phi$ where
$i\hbar\langle\xi_n|\partial_{t}\xi_n\rangle=0$.

The off-diagonal term of ${\mathbf A}$ proportional to
$\Hat{F}_x~(\Hat{F}_y)$ arises from the gradient of $|\xi_n\rangle$
in the phase $\alpha$ (amplitude depending on $\beta$). Since our
dressed atoms are prepared in $|\xi_0\rangle$ and
$\Delta\ell=\hbar$, we consider $n=0,n^{'}=-1$,
\begin{align}\label{eq:H_offdiagonal}
H_{-1,0}= -\frac{L_z\hbar}{\sqrt{2}mr^2}\sin{\beta}
+\frac{\hbar^2}{2\sqrt{2}mr^2}\sin{\beta}\cos{\beta}- \frac{\hbar
^2}{2\sqrt{2}m} \partial_r (\partial_{r}\beta) -\frac{\hbar
^2}{2\sqrt{2}m}\left(\partial_{r}\beta\right) \partial_r-\frac{\hbar
^2}{2\sqrt{2}mr}\partial_{r}\beta
\end{align}
and in Eq.~(2),
\begin{align}\label{eq:H_diagonal}
H_{\rm eff}^{(n)}= -\frac {\hbar^2}{2m}
\nabla^2(r,z)+\frac{L_z^2}{2mr^2}+\frac{L_zn\hbar}{mr^2}\cos{\beta}
+\frac{\hbar^2}{2m} \langle\nabla \xi_n| \nabla
\xi_n\rangle+\varepsilon_n+V(r).
\end{align}
In Eq.~\eqref{eq:H_diagonal}, the third term corresponds to the
cross terms of $L_z$ and $r A_n$ in Eq.~(2), and the fourth term is
given by
\begin{align}
\frac{(rA_n)^2}{2mr^2}+W_n=\frac{\hbar^2}{2m} \langle\nabla \xi_n|
\nabla \xi_n\rangle,
\end{align}, where
\begin{subequations}
\begin{align}
W_n &=\frac{\hbar^2}{2m}(\langle\nabla \xi_n| \nabla \xi_n\rangle
-i^2\langle \xi_n|\partial_{\phi} \xi_n \rangle^2 )\\
W_1=W_{-1}&=\frac{1}{4m}\left[
\frac{\hbar^2}{r^2}\sin^2{\beta}+(\hbar
\partial_r \beta)^2\right]\\
W_0&=2 W_1
\end{align}
\end{subequations}
and
\begin{subequations}
    \begin{align}
    \frac{\hbar^2}{2m}\nabla\xi_1^{\dag}\cdot \nabla \xi_1
    =\frac{\hbar^2}{2m}\nabla\xi_{-1}^{\dag}\cdot \nabla \xi_{-1}&=\frac{1}{4m}\left[
\frac{\hbar^2}{r^2}(1+\cos^2{\beta})+(\hbar
\partial_r \beta)^2 \right] \\
    \frac{\hbar^2}{2m}\nabla\xi_0^{\dag}\cdot \nabla \xi_0&=\frac{1}{2m}\left[
\frac{\hbar^2}{r^2}\sin^2{\beta}+(\hbar
\partial_r \beta)^2 \right]
    \end{align}
\end{subequations}

Here we show the validity of local dressed states
$|\xi_n(\vec{r},t)\rangle$ as the approximated eigenstates, and of
the adiabatic condition. Consider the energies associated with the
spatial gradient in both $\psi$ and $|\xi\rangle$; they correspond
to terms in Eq.~\eqref{eq:H_offdiagonal0}, leading to
Eq.~\eqref{eq:H_offdiagonal}. When these spatial gradient energies
are sufficiently smaller than the energy gap $\Omega_{\rm eff}(r)$,
it gives $|H_{n^{'}n}| \ll |\varepsilon_n -
\varepsilon_{n^{'}}|=\hbar\Omega_{\rm eff}$ for $|n^{'}-n|=1$. That
is, coupling between dressed states are negligible, and the
adiabatic condition is fulfilled; the eigenstates of the Hamiltonian
$\hat{H}_{\rm lab}$ are well approximated by
$|\xi_n(\vec{r},t)\rangle$, which are the eigenstates of
$\vec{\Omega}_{\rm eff}\cdot \vec{F}$. For our experiment of
$|\xi_0\rangle$ with $\ell=0$, the computed $|H_{-1,0}(r)|$ is
smaller than $(\hbar^2/\sqrt{2}mr^2)\sin{\beta}\cos{\beta}$ for all
$r$. We find $|H_{-1,0}(r)|$ is smaller than $\hbar\Omega_{\rm
eff}(r)$ at $r\gtrsim 0.6~\mu$m where the transition from $n=0$ to
$n^{'}=-1$ is negligible.

As we include the quadratic Zeeman energy $\hbar\omega_q
\Hat{F}_z^2$ with $\omega_q/2\pi=50$~Hz, it adds an offset to
$(\hbar^2/2mr^2)\Hat{F}_z^2$. The off-diagonal coupling $H_{-1,0}$
has an additional term $\hbar
\omega_q\sin{\beta}\cos{\beta}/\sqrt{2}$, which is smaller than
either $(\hbar^2/2\sqrt{2}mr^2)\sin{\beta}\cos{\beta}$ or
$\hbar\Omega_{\rm eff}(r)$, and thus the effects from $\hbar
\omega_q \Hat{F}_z^2$ are negligible.

\subsection{Adiabaticity of the loaded dressed state}
Our atoms are loaded into the eigenstate well approximated by the
local dressed state $|\xi_{0}\rangle$ for $r > r_c$, where the
adiabatic condition is fulfilled and $r_c$ is the adiabatic radius.
Using TDGPE simulations, we obtain the state after the loading,
$\psi(\vec{r})|\xi(\vec{r})\rangle$. At $\delta=0$, we compute the
overlap of $|\xi(\vec{r})\rangle$ to the local dressed state
$|\xi_0(\vec{r})\rangle$, where the projection probability $|\langle
\xi_0(\vec{r})|\xi(\vec{r})\rangle|^2$ exceeds $0.98$ at $r>r_c
\approx 1.4~\mu$m. The energy gap $\Omega_{\rm eff}(r)$ is
sufficiently large for the small loading speed $\dot{\delta}$ and
small spatial gradient energies for $r>r_c$, where the adiabatic
condition holds.

\subsection{Local dressed states with mean field interactions}
The projected Hamiltonian $H_{\rm eff}^{(n)}$ in Eq.~(2) is for
non-interacting atoms with $\omega_q=0$. Here we include the mean
field interactions $\int d^3\vec{r}n(\vec{r})\frac{1}{2}\left[c_0
n(\vec{r})+c_2 n(\vec{r}) \langle \vec{F} \rangle ^2\right]$ in
$F=1$ BECs, $n=n_1+n_0+n_{-1}$ is the total density and $n_{m_F}$ is
the density of bare spin state $|m_F\rangle$. The coupled spinor
TDGPE for $\Omega_{\rm eff}=0$ and $\omega_q=0$ is
\begin{subequations}\label{eq:spinorGPE}
\begin{align}
i \hbar \frac{\partial \Psi_{1}}{\partial t}&= h_0 \Psi_{1}+c_0
n\Psi_{1}+c_2 (n_1+n_0-n_{-1})\Psi_{1}+c_2
\Psi_{-1}^{*}\Psi_{0}\Psi_{0},\\
i \hbar \frac{\partial \Psi_{0}}{\partial t}&= h_0 \Psi_{0}+c_0
n\Psi_{0}+c_2 (n_1+n_{-1})\Psi_{0}+2 c_2
\Psi_{0}^{*}\Psi_{1}\Psi_{-1},\\
i \hbar \frac{\partial \Psi_{-1}}{\partial t}&= h_0 \Psi_{-1}+c_0
n\Psi_{-1}+c_2 (n_{-1}+n_0-n_{1})\Psi_{-1}+c_2
\Psi_{1}^{*}\Psi_{0}\Psi_{0},
\end{align}
\end{subequations}
where $\Psi_{m_F}=\langle m_F|\Psi\rangle$, $h_0=-(\hbar^2 /2m)
\nabla^2+V(r)$, $n_{m_F}=|\Psi_{m_F}|^2$. Since the bare spin
$|m_F\rangle$ is mapped to the dressed spin $|\xi_{m_F}\rangle$ by a
local spin rotation $\mathcal{U}$, $|\langle \vec{F}\rangle|^2$ is
invariant with respect to $\mathcal{U}$ and
$|\xi_{m_{F}}(\vec{r},t)\rangle$ remains the eigenstates when we
include the mean field.

\subsection{Time evolutions}
We discuss the stability of the prepared dressed atoms with the
Raman fields on for a hold time $t_h>0$. Without the interaction
$H_{\rm int}$ and neglecting $\omega_q$, the total Hamiltonian after
the local spin rotation for removing the dependence of
$\vec{\Omega}_{\rm eff}\cdot \vec{F}$ on $\phi$ is
\begin{align*}
\hat{H_1}=\left[\frac{-\hbar^2}{2m}\nabla^2(r,z) +\frac{L_z^2}{2m
r^2}+V(r) \right] \otimes {\hat 1} +\hbar \delta \Hat{F}_z+\hbar
\Omega \Hat{F}_x+ \hat{H}_{\rm
SOAMC}+\frac{\hbar^2}{2mr^2}\Hat{F}_z^2.
\end{align*}
When $\partial_r$ is neglected, the eigenstates are the ``modified
local dressed states" $|\bar{\xi}_n(\ell,r)\rangle$ at fixed $r$
with the good quantum number $\ell$, and the wave function in the
$n-$th dressed state in position representation is
\begin{align}\label{eq:Total_state}
\langle
\vec{r}|\Psi\rangle&=\varphi_n(\ell,r,z)\langle \phi|\bar{\xi}_n(\ell,r)\rangle\nonumber\\
&=\varphi_n(\ell,r,z)e^{i \ell \phi}U(\ell,r)|n\rangle,
\end{align}
and $|n\rangle$ is the bare spin state. At $\ell=0$,
$|\bar{\xi}_n(\ell,r)\rangle \approx |\xi_n\rangle$ at $r\gtrsim
0.6~\mu$m where the effects from $(\hbar^2/2mr^2)\Hat{F}_z^2$ are
negligible. At large $\ell$, $|\bar{\xi}_n\rangle$ deviates from
$|\xi_n\rangle$ owing to the $\hat{H}_{\rm SOAMC}=(\hbar/mr^2)\ell
\Hat{F}_z$.

Our dressed atoms are prepared in $|\xi_0\rangle \approx
|\bar{\xi}_0\rangle$ at $\ell=0$, where the final probability
projected to $|\xi_0\rangle$ or $|\bar{\xi}_0\rangle$ is close to 1.
Consider single-atom induced coupling from the initial $n=0$ state
to $n=-1$ ground dressed state. We use 2D TDGPE to simulate the
state after a hold time $t_h=0.1$~s, where it shows atoms decay to
$|\bar{\xi}_{-1}\rangle$ within $|\delta|/2\pi\lesssim 0.8~$~kHz.
Here $\ell=0$ remains unchanged, and we use initial states with
off-centered vortex position to simulate the experiment with
pointing stabilities of the laser beams. This decay is most likely
due to terms with $\partial_r$ in $\hat{H}_1$, where the initial
state of the dressed atoms is close to $|\bar{\xi}_0\rangle$. If we
consider the initial state as $|\xi_0\rangle$, the coupling to
$|\xi_{-1}\rangle$ would be given by Eq.~\eqref{eq:H_offdiagonal}.
The actual dynamics is dictated by the prepared dressed state at
$t_h=0$ and the following evolution from the TDGPE.
%there is insufficient information in whether the prepared dressed
%state is in the exact eigenstate of $\hat{H}_1$ at large $r$ where
%adiabatic loading holds. Thus it is not granted to apply
%time-dependent perturbation to derive the coupling matrix element.

Next, beyond the mean field description, we consider interactions in
the second quantization form with the leading term from $c_0$,
\begin{align*}
\hat{H}_{\rm int}^{c_0}=\frac{c_0}{2}\int d^3 \vec{r}
\sum\limits_{\sigma_A,\sigma_B}
\hat{\psi}^{\dag}_{\sigma_A}(\vec{r})
\hat{\psi}^{\dag}_{\sigma_B}(\vec{r})
\hat{\psi}_{\sigma_A}(\vec{r})\hat{\psi}_{\sigma_B}(\vec{r}).
\end{align*}

After a Fourier transform along $\phi$ it becomes
\begin{align}
\hat{H}_{\rm int}^{c_0}=2\pi \frac{c_0}{2} \int dz\int dr r
&\sum\limits_{\ell_1,\ell_2,\ell_3,\ell_4}\sum\limits_{\sigma_A,\sigma_B}
\delta_{\ell_1+\ell_2,\ell_3+\ell_4} \hat{\phi}^{\dag}_{
\sigma_B}(\ell_4)\hat{\phi}^{\dag}_{\sigma_A}(\ell_3)\hat{\phi}_{\sigma_B}(\ell_2)\hat{\phi}_{\sigma_A}(\ell_1),
\end{align}
where $\hat{\phi}_{\sigma}(\ell)$ is the operator annihilating one
atom with $\ell$ in bare spin $|m_F=\sigma\rangle$ at $(r,z)$.
$\hat{H}_{\rm int}^{c_0}$ can couple two atoms in
$|\bar{\xi}_{0}\rangle$ to the ground dressed state
$|\bar{\xi}_{-1}\rangle$ with $\ell\neq 0$, where the energy of the
initial two-atom state $|i\rangle$ matches that of the final state
$|f\rangle$. Due to the nonzero $\ell$ acquired,
$|\bar{\xi}_{-1}\rangle$ is the relevant dressed state, instead of
$|\xi_{-1}\rangle$. The resonant coupling gives a decay rate of
atoms prepared in $|\bar{\xi}_{0}\rangle$ from Fermi's golden rule
(FGR). For coupling two atoms in $n=0$ to $n=-1$, we transform
$H_{\rm int}^{c_0}$ to the field operators in the dressed spin
basis,
\begin{align}
\hat{H}_{\rm int}^{c_0}=2\pi \frac{c_0}{2} \int dz\int dr r
&\sum\limits_{\ell_1,\ell_2,\ell_3,\ell_4}\delta_{\ell_1+\ell_2,\ell_3+\ell_4}
\hat{\varphi}^{\dag}_{-1}(\ell_4)\hat{\varphi}_{0}(\ell_2)\hat{\varphi}^{\dag}_{-1}(\ell_3)\hat{\varphi}_{0}(\ell_1)\nonumber\\
&\sum\limits_{\sigma_A,\sigma_B} U^{\dag}_{-1\sigma_B}(\ell_4)
U_{\sigma_B 0}(\ell_2) U^{\dag}_{-1\sigma_A}(\ell_3) U_{\sigma_A
0}(\ell_1),
\end{align}
where $\hat{\varphi}_{n}(\ell)$ is the operator annihilating one
atom with $\ell$ in dressed spin $|\bar{\xi}_{n}\rangle$ at $(r,z)$,
\begin{align*}
\hat{\varphi}_{n}(\ell)=\hat{\varphi}_{n}(\ell,r,z),U(\ell)=U(\ell,r).
\end{align*}

Given that $\ell_1=\ell_2=0$ for $n=0$, $\ell_3(\ell_4)=+(-)\ell_f$
for $n=-1$ owing to $\ell_3(\ell_4)=+(-)\ell_f+\ell_{\rm
min}^{(-1)}\approx+(-)\ell_f$ for $|\ell_{\rm min}^{(-1)}| <\hbar
\ll \ell_f$, and $\tilde{r}\equiv(r,z)$,
\begin{subequations}\label{eq:ifstates}
\begin{align}
|i\rangle &= \int
d\tilde{r_1}d\tilde{r_2}\varphi_0(\ell=0,\tilde{r_1})
\varphi_0(\ell=0,\tilde{r_2})\frac{1}{\sqrt{2}}\hat{\varphi_0}^{\dag}(\ell=0,\tilde{r_2})\hat{\varphi_0}^{\dag}(\ell=0,\tilde{r_1})|0\rangle,\\
|f\rangle &= \int d\tilde{r_3}d\tilde{r_4}
\frac{1}{\sqrt{2}}\left[\varphi_{-1}(\ell_f,\tilde{r_3})
\varphi_{-1}(-\ell_f,\tilde{r_4})+\varphi_{-1}(\ell_f,\tilde{r_4})
\varphi_{-1}(-\ell_f,\tilde{r_3})\right]\frac{1}{\sqrt{2}}\hat{\varphi_{-1}}^{\dag}(-\ell_f,\tilde{r_4})\hat{\varphi_{-1}}^{\dag}(\ell_f,\tilde{r_3})|0\rangle,
\end{align}
\end{subequations}
where $\varphi_0(0,\tilde{r}),\varphi_{-1}(\pm \ell_f,\tilde{r})$
are normalized single particle wave functions,
\begin{align*}
\int dz\int dr 2\pi r|\varphi_n(\ell,r,z)|^2=1.
\end{align*}
This leads to the matrix element at $\ell_f$
\begin{align}\label{eq:matrix_off_lf}
\langle f|H_{\rm int}^{c_0}|i\rangle_{\ell_f}= 2\pi\cdot
2\sqrt{2}\cdot\frac{c_0}{2} \int_{-z_r}^{z_r}
dz\int_{\bar{r}_c}^{R_{\rm TF}} dr r
&{\varphi}^{*}_{-1}(-\ell_f,r,z){\varphi}^{*}_{-1}(\ell_f,r,z){\varphi}_{0}(0,r,z){\varphi}_{0}(0,r,z)\nonumber\\
&\left[U^{\dag}(-\ell_f,r)U(0,r)\right]_{-1,0}\left[U^{\dag}(\ell_f,r)
U(0,r)\right]_{-1,0}.
\end{align}
Here $z_r=\sqrt{{R_{\rm TF}}^2-r^2}$, $\bar{r}_c$ indicates that the
atoms are prepared in $|\bar{\xi}_{0}\rangle$ with the probability
exceeding $\bar{p}_0$ at $r>\bar{r}_c(\delta)$; $\bar{r}_c$ deviates
slightly from $r_c$ for loading into $|\xi_0\rangle$. At small
detuning and $r<\bar{r}_c$ it is invalid to take the initial state
as $|\bar{\xi}_0\rangle$, thus we cannot apply FGR. Here we use
$\bar{p}_0=0.9$ to determine $\bar{r}_c(\delta)$. The motional wave
functions of $|f\rangle$ are $\varphi_{-1}(\pm\ell_f)$, and
\begin{align}\label{eq:psin1_eq}
\left[ -\frac{\hbar^2}{2m}\nabla^2(r,z)+V(r,z)+c_0 n_{\rm
BEC}(r,z)+\bar{\varepsilon}(\ell_f,r)-\mu
\right]\varphi_{-1}(\ell_f) =\lambda_{E}\varphi_{-1}(\ell_f)
\end{align}
with the eigenenergy $\lambda_{E}$ which is closest to zero, since
we consider the near-resonant coupling of $|i\rangle$ to
$|f\rangle$. $\bar{\varepsilon}_{-1}(\ell_f,r)$ is the eigenenergy
of $|\bar{\xi}_{-1}\rangle$; for $\ell_f\geq 2\hbar$,
\begin{align*}
\bar{\varepsilon}_{-1}(\ell_f,r)\approx \frac{\left[\ell_f-\ell_{\rm
min}^{(-1)}\right]^2}{2mr^2}-\sqrt{\Omega(r)^2+\delta^2}.
\end{align*}
For $\ell_f=0, \hbar$, $\bar{\varepsilon}_{-1}(\ell_f,r)$ is finite
as $r\rightarrow 0$ while the approximated form diverges. $c_0
n_{\rm BEC}(r,z)$ is the effective potential due to interactions
from most of the remaining atoms in $|\xi_0\rangle$, which is the
ground state BEC with chemical potential $\mu$. $c_0n_{\rm
BEC}=\mu-V(r,z)$ at $r<R_{\rm TF}$ and $c_0n_{\rm BEC}=0$ at
$r<R_{\rm TF}$. We define an effective potential
\begin{align}
V_{\rm eff}(r,z)=[V(r,z)-\mu]\theta(\sqrt{r^2+z^2}-R_{\rm
TF})+\frac{\left[\ell_f-\ell_{\rm
min}^{(-1)}\right]^2}{2mr^2}-\sqrt{\Omega(r)^2+\delta^2},
\end{align}
$\theta$ is the Heaviside step function, and
\begin{align}
\left[ -\frac{\hbar^2}{2m}\nabla^2(r,z)+V_{\rm
eff}\right]\varphi_{-1}(\ell_f)=\lambda_{E}\varphi_{-1}(\ell_f).
\end{align}
This shows the dressed state energy $\Omega_{\rm
eff}=\sqrt{\Omega(r)^2+\delta^2}$ can be converted to the sum of
radial, azimuthal and axial kinetic energy. $\ell_f/\hbar \lesssim
100$ where the maximal $\ell_f$ corresponds to zero overlap between
$\varphi_{-1}(\ell_f)$ and $\varphi_{0}(0)$ due to the
$\ell_f^2/2mr^2$ barrier. $\varphi_{0}(0)$ is the ground state BEC
in $|\xi_0\rangle$ with TF profile.

The spin-dependent terms in Eq.~\eqref{eq:matrix_off_lf} are
\begin{align}\label{eq:spin_matrix_off_exact}
\left[U^{\dag}(-\ell_f) U(0)\right]_{-1,0}\left[U^{\dag}(\ell_f)
U(0)\right]_{-1,0}=\langle
\bar{\xi}_{-1}(-\ell_f)|\bar{\xi}_0(\ell=0) \rangle \langle
\bar{\xi}_{-1}(\ell_f)|\bar{\xi}_0(\ell=0) \rangle.
\end{align}
This indicates the spin parts of $|\bar{\xi}_0(\ell=0) \rangle$ and
$|\bar{\xi}_{-1}(\ell_f) \rangle$ are non-orthogonal, leading to the
spin decay due to collisions under SOAMC. As we neglect
$(\hbar^2/2mr^2)\Hat{F}_z^2$ at large $r>r_c$ and take $\hat{H}_{\rm
SOAMC}$ as an effective detuning $\pm \hbar\ell_f/mr^2$, $U(\ell)$
corresponds to an Euler rotation. It is $R_y(\beta)$ for
$|\bar{\xi}_0\rangle$ and is $R_y(\beta_{\pm})$ for
$|\bar{\xi}_{-1}(\pm \ell_f)\rangle$, where $\beta_{\pm}$
corresponds to $\ell=\pm \ell_f$ of atoms in
$|\bar{\xi}_{-1}\rangle$. Thus
\begin{align*}
\sum\limits_{\sigma} U^{\dag}_{-1\sigma}(\pm\ell_f) U_{\sigma
0}(0)=\langle
-1|R_y^{\dag}(\beta_{\pm})R_y(\beta)|0\rangle=\frac{\sin
(\beta_{\pm}-\beta)}{\sqrt{2}},
\end{align*}
given by the off-diagonal matrix elements of
$R_y^{\dag}(\beta_{\pm})R_y(\beta)$, and
\begin{align*}
\tan
\beta_{\pm}(r,\ell_f,\delta)=\frac{\hbar\Omega/E_L}{\hbar\delta/E_L
\pm 2\ell_f/\Delta\ell},
\end{align*}
where $E_L(r)={\Delta\ell}^2/2mr^2=\hbar^2/2mr^2$. At large $r$ and
large $\Omega_{\rm eff}$, $|\beta_{\pm}-\beta|$ is small, leading to
\begin{align}\label{eq:spin_matrix_off}
\left[U^{\dag}(\pm\ell_f) U(0)\right]_{-1,0}\approx
\sqrt{2}\frac{\ell_f}{\hbar}\frac{E_L(r)}{\hbar \Omega_{\rm
eff}}\frac{\Omega}{\Omega_{\rm eff}}.
\end{align}

Now we compute $\langle f|H_{\rm int}^{c_0} |i\rangle_{\ell_f}$ from
Eq.~\eqref{eq:matrix_off_lf} and
Eq.~\eqref{eq:spin_matrix_off_exact}. It is an overlap integral
containing $\varphi_0$ for $\sqrt{r^2+z^2}<R_{\rm TF}$, within which
the inner classical turning point of $\varphi_{-1}(\ell_f)$ is a
$z$-independent $r_{\rm min}(\ell,\delta)$. The classically
accessible region with $|\varphi_0|^2>0$ is bounded by $r_{\rm
min}<r<\sqrt{R_{\rm TF}^2-z^2}$ at given $z$, and $|z|<\sqrt{R_{\rm
TF}^2-r_{\rm min}^2}$. It is in the WKB regime with short
wave-length $\lambda_{\rm WKB}$ and slowly varying potential, thus
we only compute within the classically accessible region. In the
classically forbidden region $\varphi_{-1}(\ell_f)$ exponentially
decays within a short length scale $\approx \lambda_{\rm WKB}$, and
thus neglected. Without numerically solving $\varphi_{-1}$, we
approximate Eq.~\eqref{eq:matrix_off_lf} by using a dimensional
analysis: we take the typical single-atom 3D density of
$|\varphi_{-1}|^2$ as $\sqrt{2}/(R_z \pi(r_{\rm max}^2-r_{\rm
min}^2))$, where $R_z$ is the typical radius along $z$, $r_{\rm
max},r_{\rm min}$ are the outer and inner turning points of $V_{\rm
eff}(r,z=0)$, respectively; the sizes are numerically calculated as
a function of $(\ell_f,\delta)$.

The FGR is
\begin{align}
\Gamma = \frac{2\pi}{\hbar} \sum_{\ell_f} \left|\langle f|H_{\rm
int}^{c_0} |i\rangle_{\ell_f}\right|^2 g(E,\ell_f,\delta)N^2,
\end{align}
where $g(E)$ is the density of state in a trap at energy
$E=\lambda_E$ (see Eq.~\eqref{eq:psin1_eq}),
\begin{align*}
g(E,\ell_f,\delta)=2\pi\frac{(2m)^{3/2}}{h^3}\int
d^3\vec{r}\sqrt{E-V_{\rm eff}(r,z)},
\end{align*}
and $N^2$ factor appears since we used normalized single particle
wave functions $\varphi_0,\varphi_{-1}$. Applying the spin-coupling
terms in Eq.~\eqref{eq:spin_matrix_off} to
Eq.~\eqref{eq:matrix_off_lf}, and for $g(E,\ell_f)$ we make an
estimate without integrating within the volume of classically
accessible region: We integrate $2\pi r\sqrt{E-V_{\rm eff}(r,z)}$
within $r_{\rm min}<r<r_{\rm max}$ at $z=0$, and then times $2 \cdot
R_z$ without integrating along $z$. We found $g(E,\ell,\delta)$
insensitive to $(\ell,\delta)$.

\begin{figure}
    \centering
    \includegraphics[width=3.4 in]{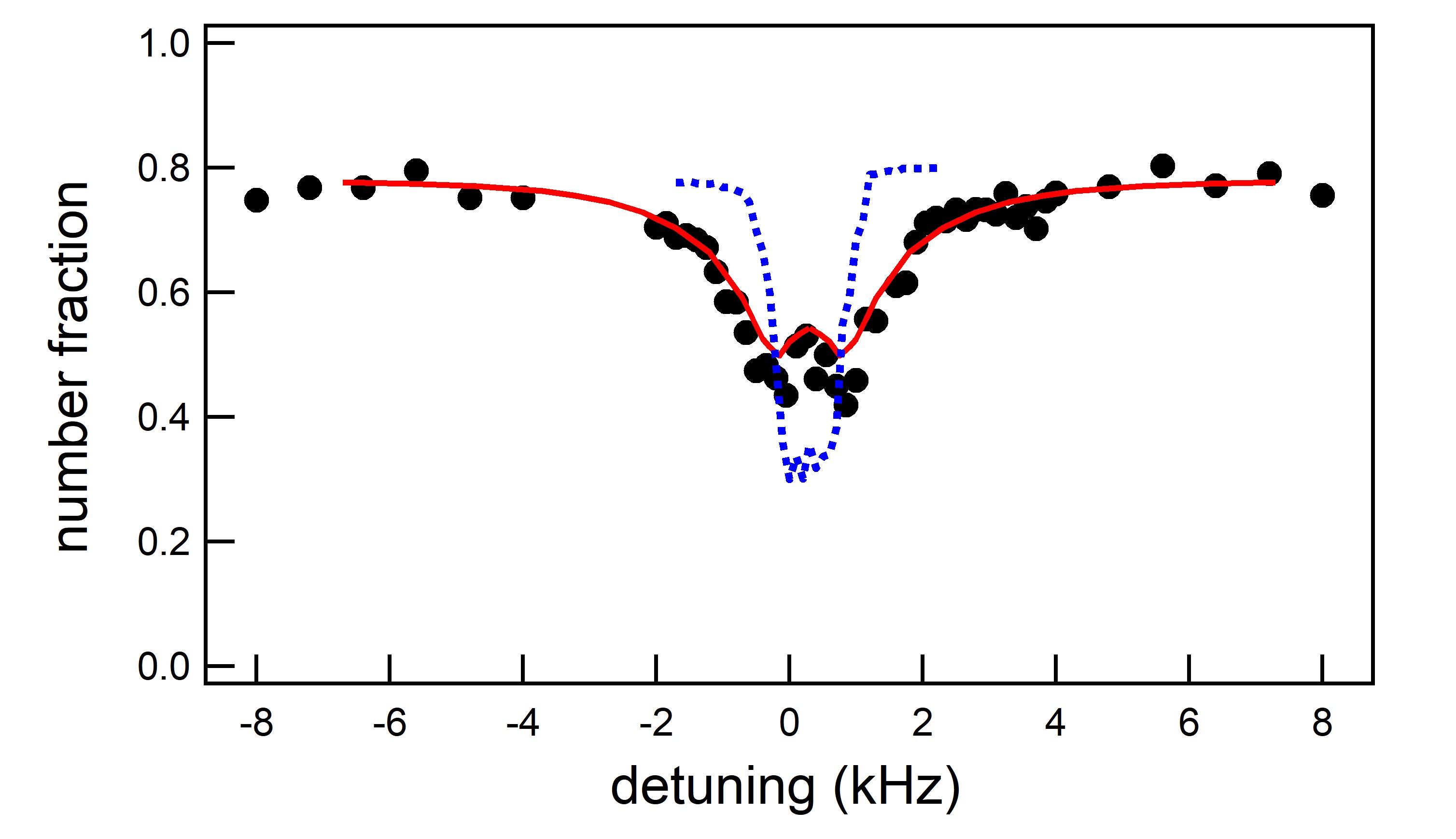}

        \caption
    {Number fraction in the dressed state $|\xi_0\rangle$ after a $t_h=0.1$~s hold
    time with Raman fields on versus detuning $\delta$ (symbols); the fraction is normalized to the total
    number at $t_h=0.1$~s. The red curve denotes calculated loss from collision-induced decay with a correction factor.
    The dotted curve denotes the 2D TDGPE simulation with the vortex position off-centered by $0.5~\mu$m in the initial state.
    Both simulation curves include a technical loss rate, see text.
    }
\end{figure}

\section{Data and simulations for the decay of dressed states}

For the data of decay of dressed state $|\xi_0\rangle$ in Fig.~3, we
display the fraction $f_{0}$ of the atom number in
$|\bar{\xi}_{0}\rangle$ over the total number in
$|\bar{\xi}_{0}\rangle$ and $|\bar{\xi}_{-1}\rangle$ at $t_h=0.1$~s.
With this normalization, the fraction would remain 1.0 with a finite
one-body loss rate from spontaneous photon scattering in the Raman
beams. Atoms initially in $|\bar{\xi}_0(\ell=0)$ are coupled to the
energy-matched states in the ground dressed state
$|\bar{\xi}_{-1}(\ell=\pm \ell_f)\rangle$ with $\ell_f > 0$, since
the spin parts of $|\bar{\xi}_0(\ell=0) \rangle$ and
$|\bar{\xi}_{-1}(\ell_f) \rangle$ are non-orthogonal. The fraction
$f_{0}$ decays faster with decreasing $|\delta|$; near the resonance
$\delta=0$, the lifetime reaches the minimum of $0.1$~s. For
$|\delta/2\pi|$ exceeding 8~kHz, $f_{0}(t_h=0.1~\rm s)$ reaches
$\approx 80\%$ instead of unity. This disagreement comes from
experimental imperfections and can be improved with better Raman
beam alignment: the loss is from the retro-reflection of one Raman
beam together with the other Raman beam, which drive two-photon
transitions resonant at $\delta/2\pi\sim \pm 14$~kHz, about four
times the photon recoil energy.

To compare the data with simulations, we first evaluate $\Gamma/N$
as described earlier, and multiply a correction scaling factor 4.2;
such factor is expected given that we have used dimensional analysis
for $|\varphi_{-1}|^2$ in Eq.~\eqref{eq:matrix_off_lf}. Another
possibility is that the collision may have other channels such as
$|\bar{\xi}_0\rangle \bigotimes |\bar{\xi}_0\rangle \rightarrow
|\bar{\xi}_1\rangle\bigotimes|\bar{\xi}_{-1}\rangle$ with the same
order of magnitude. $\Gamma/N$ is also the transition rate per atom
$dN/dt/N$ at $t_h=0$. We compare the obtained fraction
$\exp[-(\Gamma/N+\gamma_0)t_h]$ at $t_h=0.1$~s with the experimental
data.  Here, $\gamma_0=2.5 ~{\rm s}^{-1}$ accounts for the technical
loss rate corresponding to the $\approx 0.8$ fraction for
$t_h=0.1$~s at large detunings (see the previous paragraph).  The
computed fraction is shown with the data in Fig.~S2. We also compare
the results with that of 2D TDGPE simulations with the vortex
position off-centered by a typical value of $0.5~\mu$m in the
initial state for our experiment. This simulation shows losses
within a smaller detuning range of $|\delta|\lesssim 0.8$~kHz than
the data, and the curve is insensitive to the amount of off-centered
vortex position between $0.2~\mu$m to $1.0~\mu$m. We obtain the same
detuning range for a centered vortex with a small spatially random
noise in the initial state. The simulation for a vortex off-centered
by $0.5~\mu$m is displayed in Fig.~S2 after being multiplied by a
factor of $\exp(-\gamma_0t_h)=0.8$. For simplicity we do not include
the loss from 2D TDGPE in Fig.~3

For the simulation of $\Gamma/N$, we find at $\delta/2\pi \gtrsim
1$~kHz, the calculated rate $\Gamma(\delta)$ is insensitive to the
choice of $\bar{p}_0$. While at $\delta/2\pi \lesssim 1$~kHz, the
rate is notably larger with smaller $\bar{r}_c(\delta)$ set by a
smaller $\bar{p}_0$. Thus, we find the approach of time-dependent
perturbation and FGR are valid at $\delta/2\pi \gtrsim 1$~kHz.

\section{Notes on SOAMC systems}
\subsection{Cylindrical symmetry of SOAMC}
In the comparison of SOAMC and SLMC with the specific case of
effective rotations, or synthetic magnetic fields, leading to
vortices in the ground state BEC, both schemes can achieve it.
However, SOAMC can do this in a way that is cylindrically symmetric
while SLMC cannot. One result of this difference is that for the
lowest energy dressed state, SOAMC and SLMC give an anti-trapping
potential along ${\mathbf e}_{r}$ and ${\mathbf e}_{y}$,
respectively, due to the position dependent energy eigenvalues,
$-\sqrt{\Omega_{\rm eff}^2+\delta^2}$ where $\Omega_{\rm
eff}=|\vec{\Omega}_{\rm eff}|$.

To continue the above discussions, we compare SOAMC and SLMC with an
identical synthetic magnetic field $\vec{B}^*= \nabla \times
\vec{A}$, which is uniform along ${\mathbf e}_{z}$: In SOAMC,
dressed eigenstates have angular momentum as the good quantum
number, which doesn’t hold for SLMC. This makes the wave functions
in these two gauges have different phases, which is revealed in the
measurement process when the synthetic gauge field is turned off.
(see Ref.~S\cite{LeBlanc2015s}) In the usually adopted TOF method,
the cloud expands symmetrically in the gauge of SOAMC, but not in
that of SLMC. Here, the phase winding of the cloud is the same in
both gauges although the expansion is different.

We now discuss topological spin excitations created by SOAMC.
Topological spin textures with cylindrical symmetry, such as
coreless vortices, skyrmions and monopoles, can not be achieved with
SLMC, as we explain in the following. We consider dressed states in
the spin $|<\vec{F}>|$=1 manifold, i.e., $<\vec{F}>$ aligns with
$\vec{\Omega}_{\rm eff}$. Both spin textures of the coreless vortex
and monopole have cylindrical symmetry and the direction of spin
$<\vec{F}>$ winds by $2\pi$ as $\phi$ varies from 0 to $2\pi$. In
SLMC, the spin projected on $xy$ plane has a helically precessing
angle $2k_rx$ along ${\mathbf e}_{x}$, and it cannot be consistent
with spin textures with cylindrical symmetry. We list examples for
above statements in three cases, where we explicitly show the form
of unit vector of $\vec{\Omega}_{\rm eff}$, along which the local
spin aligns.  (i) ${\mathbf e}_{\rm v}$ for a coreless vortex in
SOAMC (ii) ${\mathbf e}_{\rm m}$ for a monopole generated by spin
rotations with real magnetic fields (iii) ${\mathbf e}_{\rm SLMC}$
for the SLMC. We have
\begin{align*}
{\mathbf e}_{\rm v}&=\sin\beta(r) \cos\phi{\mathbf
e}_{x}-\sin\beta(r) \sin\phi{\mathbf e}_{y}+\cos\beta(r){\mathbf
e}_{z},\\
{\mathbf e}_{\rm m}&=\sin\theta^{'} \cos\phi^{'}{\mathbf
e}_{x}+\sin\theta^{'} \sin\phi^{'}{\mathbf
e}_{y}-\cos\theta^{'}{\mathbf
e}_{z},\\
{\mathbf e}_{\rm SLMC}&=\sin\beta_1(y) \cos [2k_r x]{\mathbf
e}_{x}-\sin\beta_1(y) \sin [2 k_r x]{\mathbf
e}_{y}+\cos\beta_1(y){\mathbf e}_{z}.
\end{align*}
For (i), $\beta(r)=\tan^{-1}[\Omega(r)/\delta]$ and $\delta$ is
spatially uniform; for (ii), $(r^{'},\theta^{'},\phi^{'})$ is a
rescaled spherical coordinate from $(x^{'}=x,y^{'}=y,z^{'}=2z)$ (see
Ref.~\cite{Ray2014s}). For (iii),
$\beta_1(y)=\tan^{-1}[\Omega/\delta_1(y)]$ where $\Omega$ is
spatially uniform and $\hbar k_r$ is the photon recoil momentum.

Next we discuss the comparison of SOAMC to spin rotations with real
magnetic fields $\vec{B}$ for making topological excitations, and
potential studies on those with SOAMC. Since $\vec{\Omega}_{\rm
eff}$ with SOAMC can be designed with a spatial-light-modulator or
digital-mirror-device, it can have “smaller spatial scales and
faster time scales” as compared to those of spin rotations with real
magnetic fields, whose spatial scale is determined by the coil size
and time scale is limited by the coil's inductance. One obvious
advantage from small spatial scales is the capability of studying
interactions within a pair of vortex-antivortex, or a pair of
monopole-antimonopole, where probing a small pair size may be
possible. Here, we refer to monopoles (antimonopoles) as those
generated by real (light-induced) magnetic fields with $\nabla \cdot
\vec{B}=0$ ($\nabla \cdot \vec{\Omega}_{\rm eff}\neq 0$). Let's
focus on the lowest eigenenergy manifold. The synthetic magnet field
$\vec{B}^*= \nabla \times \vec{A}$ has
\begin{align*}
\vec{B}^{*}\cdot {\mathbf e}_{i}=-\frac{\hbar}{2}
\epsilon_{ijk}\hat{b}\cdot\left(\partial_j \hat{b}\times \partial_k
\hat{b}\right),
\end{align*}
where $\hat{b}$ is the unit vector of local $\vec{B}$ or
$\vec{\Omega}_{\rm eff}$. Consider a monopole with
$\hat{b}=\hat{b}_{\rm m}$ and an antimonopole with
$\hat{b}=\hat{b}_{\rm am}$, where
\begin{align*}
\hat{b}_{\rm m}=\frac{{x^{'}\mathbf e}_{x^{'}}+{y^{'}\mathbf
e}_{y^{'}}-{z^{'}\mathbf e}_{z^{'}}}{r^{'}}, \hat{b}_{\rm
am}=\frac{{x^{'}\mathbf e}_{x^{'}}+{y^{'}\mathbf
e}_{y^{'}}+{z^{'}\mathbf e}_{z^{'}}}{r^{'}}.
\end{align*}
The nonzero divergence of $\hat{b}_{\rm am}$ is made possible by
$\vec{\Omega}_{\rm eff}$ with SOAMC. One can easily check that
$\vec{B}^{*}$ for the monopole (antimonopole) is along ${\mathbf
e}_{r^{'}}$ ($-{\mathbf e}_{r^{'}}$), leading to a positive
(negative) topological charge. If we make a sign change,
$\hat{b}_{\rm m}\rightarrow -\hat{b}_{\rm m}$, and $\hat{b}_{\rm
am}\rightarrow -\hat{b}_{\rm am}$, the sign of the topological
charge changes for both the monopole and antimonopole, while the
antimonopoles always have opposite charges to that of the monopoles
which are generated by real magnetic fields.

We now discuss examples of topological excitations that can be
generated by $\vec{\Omega}_{\rm eff}$ with SOAMC, which are not
achievable in a straightforward way by spin rotations with real
magnetic fields. As mentioned previously, two examples are a pair of
monopole-antimonopole, and a pair of vortices or vortex-antivortex.
The latter can be created with two pairs of LG Raman beams; when the
propagating directions of these two pairs are not colinear, one can
study the collisions of non-colinear vortices. For instance, the
production of resulting rung vortex for non-abelian vortices in the
$F=2$ manifold can be tested.

\subsection{Proposal of measuring superfluid fractions with SOAMC}
We discuss the scheme of measuring superfluid (SF) fractions with
SOAMC in Ref.~\cite{Cooper2010s} using a spectroscopy method, i.e.,
the population imbalance of bare spin components after projection of
the dressed state. As we will show, such measurement cannot be used
to derive SF fractions with the SLMC, owing to the spin imbalance is
gauge-dependent.

In Ref.~\cite{Cooper2010s}, $F=1$ atoms are confined in a ring trap
with radius $R$ under SOAMC in the lowest energy dressed state,
which is consistent with $|\bar{\xi}_{-1}\rangle$ in our notation.
With a Raman detuning $\delta$, the effective energy dispersion of
the dressed state is $(\ell-\ell^{*})^2/2m^*R^2$, where the minimum
is at $\ell=\ell^{*}(\delta)$, corresponding to an effective
rotation and an azimuthal gauge potential. Here it is in the large
Raman coupling regime, $\hbar\Omega/E_L\gg1$,
$E_L=\Delta\ell^2/2mR^2$, and the effective mass is
$m^*=m(1+2E_L/\hbar\Omega)$. The minimum is
\begin{align}
\ell^{*}\approx \frac{\delta}{\Omega}\Delta\ell
\end{align}
for small $\delta/\Omega$. One can derive the population imbalance
between the bare spin components $|m_F=-1\rangle$ and
$|m_F=1\rangle$ as
\begin{subequations}\label{eq:SFfrac_SOAMC}
\begin{align}
|\psi_{-1}|^2-|\psi_{1}|^2&=\frac{\ell}{\Delta
\ell}-\frac{\ell-\ell^{*}}{(m^*/m)\Delta \ell}
=\Delta p_0+\Delta p^{'}\ell,\\
\Delta p_0&\approx \frac{\delta}{\Omega}\left(1- \frac{2E_L}{\hbar\Omega}\right),\\
\Delta p^{'}&\approx \frac{1}{\Delta\ell}\frac{2E_L}{\hbar\Omega}.
\end{align}
\end{subequations}
The SF has $\ell=0$ and the population imbalance $\Delta p=\Delta
p_0$; the normal fluid has $\ell=\ell^*$ with zero velocity and
$\Delta p_N \neq \Delta p_0$. To experimentally measure the SF
fraction, one needs to distinguish a SF from a normal fluid, i.e.,
to measure the $\Delta p$ with an absolute accuracy of $\Delta
p_N-\Delta p_0$.

Now we consider the SF and normal fluid under SLMC, where the Raman
coupling $\Omega_1$ is uniform and the detuning
$\delta_1(y)=\delta_1^{'} y$ has a gradient. Similar to
Eq.~\eqref{eq:SFfrac_SOAMC}, the population imbalance of the bare
spin components of the dressed state is
\begin{align}\label{eq:SFfrac_SLMC}
|\psi_{-1}|^2-|\psi_{1}|^2&=\frac{k_x}{2k_r}-\frac{k_x-k_x^*}{(m_1^*/m)2k_r},
\end{align}
where $\hbar k_x$ is the $x$ component of canonical momentum
$\vec{P}_{\rm can}$ and $k_x^{*}=-B^*y/\hbar$ is the minimum
location of the energy dispersion versus $k_x$; $B^{*}$ is the
strength of the approximately uniform synthetic magnetic field along
$z$, and $m_1^*=m(1+2E_L/\hbar\Omega_1)$. For the SF, one can
derive~\cite{LeBlanc2015s}
\begin{align}
\vec{P}_{\rm can}= -\frac{B^{*}y}{2} {\mathbf
e}_{x}-\frac{B^{*}x}{2} {\mathbf e}_{y}.
\end{align}
The SF has the spin population imbalance
\begin{align}
\Delta s_0=\frac{B^*y}{4\hbar k_r}\frac{2E_L}{\hbar \Omega_{1}}.
\end{align}
A normal fluid has the ensemble averaged $\langle k_x
\rangle-k_x^*=0$, and thus $\langle k_x\rangle=k_x^*$, leading to
\begin{align}
\Delta s_N=\frac{B^*y}{2\hbar k_r}.
\end{align}
Thus, for both the SF and normal fluid, with the spectroscopy method
the spin population imbalance is zero after being summed within the
atomic cloud.

\subsection{Practical schemes for realizing the striped phase}
The characteristic energy scale in SOAMC systems is
$E_L=\Delta\ell^2/2mR^2$, where $\Delta\ell$ is the OAM transfer
from the Raman beams and $R$ is the typical system size. For
$\Delta\ell=\hbar$ and $R=5~\mu$m, $E_L= h\times 2.3$~Hz is much
smaller than that of SLMC, $E_r\approx h\times 3.5$~kHz at
$\lambda=0.8~\mu$m.

For observing the stripe phase in SOAMC, an example with typical
experimental parameters is shown in Ref.~\cite{Qu2015s}. Here,
pseudo-spin 1/2 $^{87}$Rb BECs with Thomas-Fermi radius about
$40~\mu$m have SOAMC with two Raman LG beams carrying phase winding
numbers of $\pm 2$, and the OAM transfer is $\Delta\ell=4\hbar$. The
radius at peak intensity of both LG beams is $r_M=~17~\mu$m. At zero
Raman detuning, the critical peak Raman coupling is $h \times
0.8$~Hz for the transition between the striped phase (miscible) and
the immiscible phase; here $E_L=h\times 0.6$~Hz.

The critical Raman coupling $\hbar \Omega_c$ at the order of
$h\times 1$~Hz is not practical for experiments given that the
detuning noise arising from typical magnetic field noise of 1 mG is
$h\times 700$~Hz. Therefore, one needs to increase $E_L$ and thus
$\Omega_c$. From Ref.~\cite{Moulder2013s}, single high-order LG beam
with phase winding number of 100 can be made with a SLM. An estimate
of $\hbar \Omega_c$ for observing miscible stripe phases can be
$\gtrsim h\times 1$~kHz for $\Delta\ell=50\hbar$ and a condensate
radius of $10~\mu$m. Here we scale down the size of the condensate
and the radius at peak intensity of the Raman LG beams from those in
Ref.~\cite{Qu2015s}; the LG beams carry phase winding number of $\pm
25$, respectively. Consider the effects of detuning noise arising
from magnetic field noise, one can expect to suppress the noise to
$\sim h\times 100$ Hz, or 0.14~mG. For reaching field noise below
$h\times 100$~Hz, see Ref.~\cite{Trypogeorgos2018s}. The
corresponding detuning noise can be made much smaller than the
critical Raman coupling. Combining all these numbers it suggests
that observing miscible stripe phases in SOAMC system may be
possible. (Note that the parameters of phase winding number of 25
and the scaled-down $r_M=4.25~\mu$m corresponds to a waist of
$w=r_M/\sqrt{25/2}=1.2~\mu$m, which is close to the diffraction
limit given $\lambda=0.8~\mu$m. Thus one needs to use a high
numerical-aperture imaging objective.)

\section{Methods of the experiment}
\subsection{System preparation and probing}
We produce a $^{87}$Rb BEC of $N \approx 4 \times 10^5$ atoms in a
crossed dipole trap in $|1,-1\rangle$ with approaches similar to
those in Ref.~\cite{Lin09as}. The dipole trap contains two 1064 nm
laser beams propagating along ${\mathbf e}_{x^{'}},{\mathbf
e}_{y^{'}}=({\mathbf e}_{x}\pm{\mathbf e}_{y})/\sqrt{2}$ with beam
waists of $\sim 65~\mu$m, and the trap frequencies for the BEC are
$(\omega_{x^{'}},\omega_{y^{'}},\omega_z)/2\pi$=(72,72,81)~Hz. After
the BEC production we wait for the external trigger from the 60 Hz
line, after which we apply feed-forward current signals into bias
coils to cancel the field noise from 60 Hz harmonics (see later
discussions). Then we transfer the BEC to $|1,0\rangle$ by first
applying a microwave $\pi$ pulse at $|1,-1\rangle \rightarrow
|2,0\rangle$ transition, followed by a second $\pi$ pulse at
$|2,0\rangle \rightarrow |1,0\rangle$. We confirm there were no
discernible atoms left in $|1,-1\rangle$, and blow away the residual
$|2,0\rangle$ atoms with a resonant $F=2 \rightarrow F^{'}=3$ pulse.
We then again wait for the 60 Hz trigger, and load the atoms into
the Raman dressed state with the following procedures. We ramp the
detuning to $\delta/2\pi=5$ kHz while the Raman beams are off, ramp
$\Omega(r,t)$ in 15~ms to the final value of $\Omega_M/2\pi= 10$
kHz, and then ramp the detuning to $\delta_f/2\pi$ between 4 kHz and
-1 kHz with $\dot{\delta}/2\pi= -1.67$~kHz/ms (see Fig.~S3),
subsequently holding $\Omega_M$ and $\delta$ at constant for $t_h$.
The Raman beams are at $\lambda=790$~nm where their scalar light
shifts from the D1 and D2 lines cancel. The Gaussian Raman beam has
a waist of $200~\mu$m, and the LG Raman beam produced by a vortex
phase plate has a phase winding number $m_{\ell}=1$ and radial index
of 0. The Raman beams are linearly polarized along ${\mathbf e}_{x}$
and ${\mathbf e}_{y}$, respectively.

\begin{figure}
    \centering
    \includegraphics[width=4.9 in]{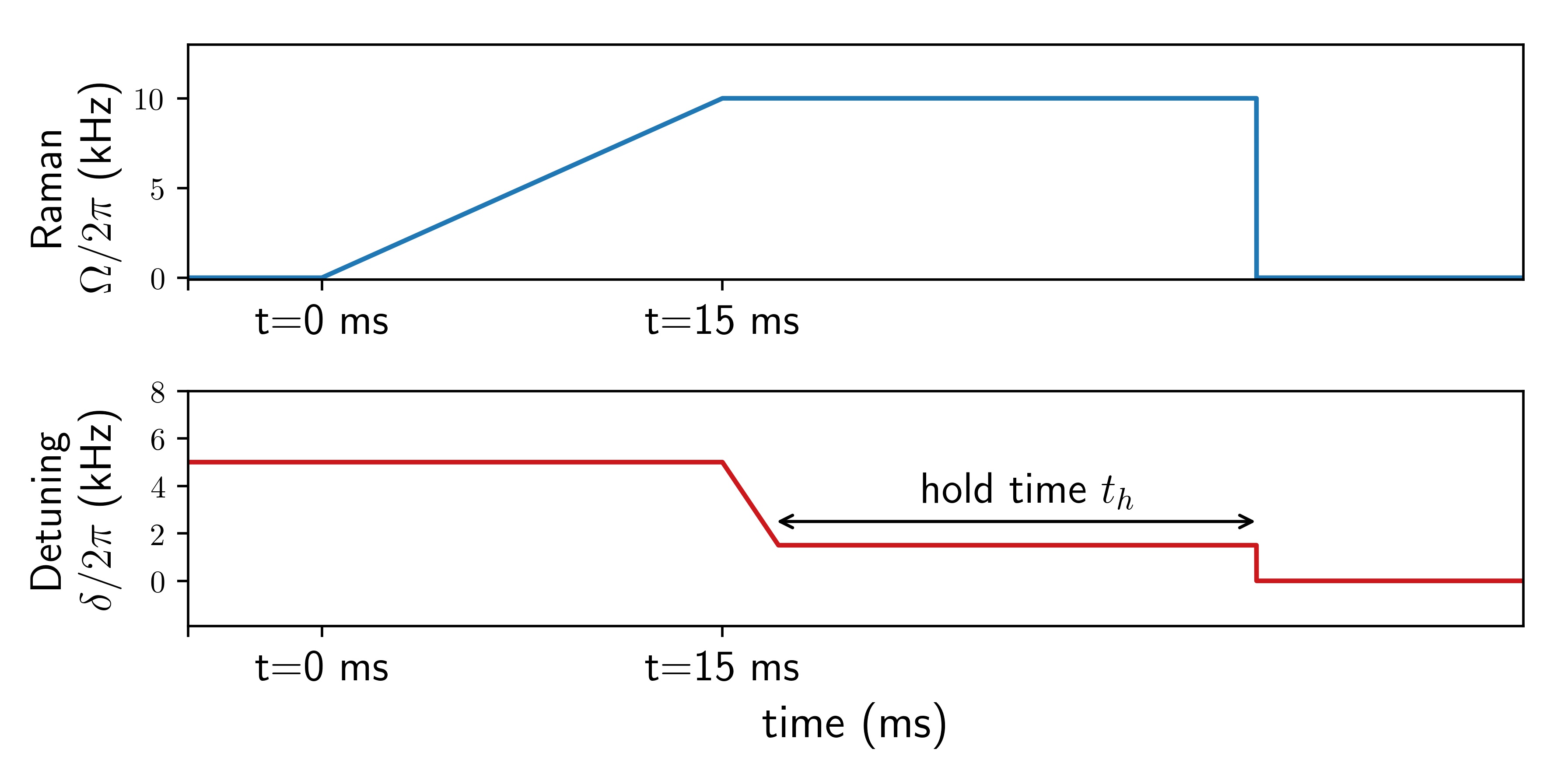}

        \caption
    {Time sequences of the Raman coupling and detuning for loading atoms
    into the dressed state.
    }
\end{figure}

For projection measurements of the spinor state $|\xi_s\rangle$, we
abruptly turn off the dipole trap and Raman beams, simultaneously
and adiabatically rotate the magnetic bias field from along
${\mathbf e}_{x}$ to that along the imaging beam direction within
$0.4$ ms; this projects $|\xi_s\rangle$ to the bare spin $m_F$
basis. The atoms then expand in free space with all $m_F$ components
together for a time-of-flight (TOF). To perform spin-selective
imaging, we apply a microwave pulse to drive the $|1,-1\rangle
\rightarrow |2,-2\rangle$ transition for imaging $m_F= -1$,
$|1,0\rangle \rightarrow |2,0\rangle$ pulse for imaging $m_F=0$, and
$|1,1\rangle \rightarrow |2,2\rangle$ for imaging $m_F= +1$,
respectively. These three frequencies are separated by $0.91$~MHz in
a field $\sim 1.3$~G along ${\mathbf e}_{z}$ for the vertical
imaging and by $0.42$~MHz in $\sim 0.6$~G along ${\mathbf e}_{y}$
for the side imaging. The resonances have separations much larger
than the microwave Rabi frequencies, which are between $4.3$ and
$16$ kHz. After the $F=1$ atoms are transferred to $F=2$, we apply a
resonant absorption imaging pulse of $\sim 14~\mu$s with $\sigma +$
polarization at the $|F=2,m_F=2\rangle \rightarrow
|F^{'}=3,m_F=3\rangle$ cycling transition. The saturation parameter
is $I/I_s= 1.60$ for the vertical imaging and $I/I_s=0.66$ for the
side imaging. We use the modified Beer-Lamberet
law~\cite{Reinaudi2007s} to derive correct optical densities.

The ambient field noise has a standard deviation ($\sigma$) $\sim
h\times 0.6$~kHz, which is dominated by the 60 Hz line signal and
its high-order harmonics. After we apply feed-forward signals in the
bias fields to cancel the dominating field noise at 60 Hz, 180 Hz,
and 300 Hz, the $1-\sigma$ residual field noise is $\sim 0.2$~kHz.
We prepare the dressed state after the 60 Hz line trigger in order
to reduce the shot-to-shot field variation with a fixed hold time
after the trigger. The measured $1-\sigma$ field noise from repeated
experimental shots is $\sim 0.11$~kHz.

\subsection{Interference for measuring relative phases}
We prepare the $\vec{F}=0$ polar dressed state at $\delta=0$, hold
$t_h=1$~ms, and then apply a radio-frequency(rf) $\pi/2$ pulse which
transforms the $|1,\ell_{1}\rangle$ component to
$(|1,\ell_{1}\rangle/2+|0,\ell_{1}\rangle/\sqrt{2}+|-1,\ell_{1}\rangle/2)^T$,
and the $|-1,\ell_{-1}\rangle$ component to
$(|1,\ell_{-1}\rangle/2-|0,\ell_{-1}\rangle/\sqrt{2}+|-1,\ell_{-1}\rangle/2)^T$.
This mixed angular momentum states $\ell_{1}$ and $\ell_{-1}$ into
each spin state for interference. After TOF we selectively probe the
$|1\rangle$ component; the nodal-line in Fig.~1c shows
$|\ell_{1}-\ell_{-1}|=2\hbar$ and the relative phase winding between
$|1\rangle$ and $|-1\rangle$ components of the dressed state is
$4\pi$. This is under the condition when the two Raman beams are
aligned to be co-propagating; when their propagating directions
deviated slightly the interference showed a fork-pattern like those
in~\cite{Choi2012as}.

\subsection{Deloading}
To measure the dressed atoms' projections onto individual dressed
bands by deloading, we reverse the loading sequence: we ramp the
detuning back to the initial value of $\delta/2\pi=5$~kHz with
$\dot{\delta}/2\pi= 1.67$~kHz/ms, turn off $\Omega$ in 15~ms, and
start TOF. Thus, for $r>r_c$ where it is adiabatic given the ramping
speed and a sufficiently large energy gap $\Omega_{\rm eff}(r)$,
atoms in the dressed bands
$|\bar{\xi}_{1}(r)\rangle,|\bar{\xi}_{0}(r)\rangle,|\bar{\xi}_{-1}(r)\rangle$
are mapped to the bare spin states
$|+1\rangle,|0\rangle,|-1\rangle$,
respectively~\cite{Williams12s,lin11as}. We apply Stern-Gerlach
gradient during TOF, and use a repumping laser to pump the atoms
from $|F=1\rangle$ to $|F=2\rangle$ before the absorption imaging.

Consider the condensate component before TOF starts. For dressed
atoms with the external part of wave function
$\varphi_{n,\ell}(r,z)$ and the normalized spinor state
$|\bar{\xi}_n(\ell,r)\rangle$ in Eq.~\eqref{eq:Total_state}, it is
mapped to the bare spin $|m_F=n,\ell+n\hbar \rangle$ with the
external wave function unchanged. This mapping is valid when the
$\delta$-dependent light shift potentials $\varepsilon_n$ of the
dressed state for $n=\pm 1$ are not so large to deform the external
wave function during the ramping of $\delta$ in deloading. After TOF
starts, if all the spin components expand together, after the
expansion each spin corresponds to a dilation of the in-situ profile
by the same factor under the approximation of neglected $c_2$, which
is verified by the TOF simulations (Fig.~S1). In the case with
Stern-Gerlach gradient which spatially separates the spin
components, the dilation does not apply while the respective number
in each dressed state are mapped to respective bare spin states.
Finally we consider the thermal component resulting from the
collisional relaxation from $|\bar{\xi}_{0}\rangle$ to
$|\bar{\xi}_{-1}\rangle$: after TOF it gives momentum distributions
of each spin component regardless whether the Stern-Gerlach gradient
is applied, provided the interaction during TOF is neglected.

\section{Data acquisition and analysis}
For data in Fig.~2 and Fig.~4, the imaging is performed with the
microwave spectroscopy selective to the bare spin $m_F$. i.e., each
$m_F$ image corresponds to an individual experimental realization.
The deloading data in Fig.~3 is taken with Stern-Gerlach gradient
during TOF, where images of all spin $|m_F\rangle$ states are taken
in a single shot (see inset).

For Fig.~2 data, we average over about 10 images taken under
identical conditions; this takes into account the shot-to-shot BEC
number variation with a standard deviation ($\sigma$) $\sim 3~\%$,
and reduces the photon shot noise. Given the short-term pointing
stability of the dipole beams and of the Raman LG beam which
determine the center of BEC and vortices, respectively, we
post-select images whose vortex positions in $|m_F=\pm 1\rangle$
with respect to the BEC center are $<0.63~\mu$m (converted from TOF
position to in-situ position). We determine $\delta/2\pi$ from the
rf-spectroscopy with an uncertainty of $\lesssim 0.1$~kHz. With the
given field noise at a fixed $t_h=1~$ms, we post-select images whose
optical density of the $|0\rangle$ component are within one
$\sigma$; this excludes data with large variation of $\delta$.

In Fig.~2a, at $\delta=0$ the measured spin texture fraction of
$|m_F=0\rangle$ $D_0/(D_1+D_0+D_{-1})$ is about 0.1 at $r=0$, which
is much smaller than the expected 1.0, since the vortices in $|\pm
1\rangle$ ideally have the optical densities $D_{-1}=D_1=0$ at
$r=0$. This is consistent with the observation that $D_0$'s $1/e^2$
radius is larger than the $\approx 90~\mu$m BEC radius (after TOF),
and much larger than the $\approx 15~\mu$m predicted by TOF
simulations (see Fig.~S1a). This disagreement is likely due to that
at exact resonance, $\delta=0$, the dressed state loading is
affected by technical noises in the Raman beams and the small
non-adiabatic spin fraction deviates from the prediction.

For data in Fig.~4, from individually taken images of $|m_F=\pm
1\rangle$, we post select those whose BEC centers are sufficiently
close before the Raman beam is turned off, in the presence of the
dipole beam's point stability. We collect 10 images from individual
experimental realizations for both $|1\rangle$ and $|-1\rangle$
($D_0$ is not discernible), take the sum of total optical density
$D_1+D_{-1}$ from the $10^2=100$ combinations, and fit them to 2D TF
profiles. We select the pair with best fit for each $t_{\rm on}$ and
display them in Fig.~4. For a given $t_{\rm on}$, we find that about
three best- fit pairs have similar magnetization images and are thus
representative, indicating such post-selection is effective.

\end{document}